\newcommand*\diff{\mathop{}\!\mathrm{d}}

\documentclass[a4paper,11pt]{article}
\pdfoutput=1 
\usepackage{jcappub} 

\usepackage[utf8]{inputenc}
\usepackage{tensor}
\usepackage{graphicx}
\usepackage{dcolumn}
\usepackage{bm}
\usepackage{comment}
\usepackage{xcolor}
\usepackage{booktabs}
\usepackage{xspace}
\usepackage{enumitem}
\usepackage{nameref}
\usepackage{hyperref}

\usepackage{braket}
\newcommand{\virgolette}[1]{``#1''}

\newcommand\mathperiod{\,.}
\newcommand\mathcomma{\,,}

\usepackage{xparse}

\NewDocumentCommand{\expval}{s m}{%
  \ensuremath{%
    \IfBooleanTF{#1}{%
      \left\langle #2 \right\rangle%
    }{%
      \langle #2 \rangle%
    }%
  }%
}

\NewDocumentCommand{\op}{s m}{%
  \ensuremath{%
    \IfBooleanTF{#1}{%
      \widehat{#2}
    }{%
      \hat{#2}
    }%
  }%
}
\NewDocumentCommand{\hcop}{s m}{%
  \ensuremath{%
    \IfBooleanTF{#1}{%
      \op*{#2}%
    }{%
      \op{#2}%
    }^{\dagger}%
  }%
}

\NewDocumentCommand{\moment}{s m}{%
  \ensuremath{%
    \Delta%
    \IfBooleanTF{#1}{%
      \left(#2\right)%
    }{%
      (#2)%
    }%
  }%
}

\NewDocumentCommand{\commutator}{s m m}{%
  \ensuremath{%
    \IfBooleanTF{#1}{%
      \left[#2, #3\right]%
    }{%
      [#2, #3]%
    }%
  }%
}

\NewDocumentCommand{\intmeasure}{o m}{%
  \ensuremath{%
    \diff%
    \IfValueT{#1}{^{#1}}%
    #2\,%
  }%
}

\NewDocumentCommand{\compconj}{s m}{%
  \ensuremath{%
    \IfBooleanTF{#1}{%
      \overline{#2}%
    }{%
      \bar{#2}%
    }%
  }%
}

\makeatletter
\let\orgdescriptionlabel\descriptionlabel
\renewcommand*{\descriptionlabel}[1]{%
  \let\orglabel\label
  \let\label\@gobble
  \phantomsection
  \edef\@currentlabel{#1\unskip}
  \let\label\orglabel
  \orgdescriptionlabel{#1}%
}
\makeatother

\newcommand{\gftfield}{\varphi}    %% GFT field
\newcommand{\gftfieldop}{\op{\varphi}}    %% GFT field operator
\newcommand{\gvariables}{g_{I}}    %% Group variables
\newcommand{\dimension}{d}         %% GFT field rank
\newcommand{\groupdomain}{G}       %%Group domain
\newcommand{\vspinrep}{\vec{\xi}}    %% Spin representation labels
\newcommand{\framefield}{\chi}     %% Frame field labels
\newcommand{\framefieldbf}{\boldsymbol{\chi}}
\newcommand{\matterfield}{\phi}    %% Matter field
\newcommand{\framefieldop}{\op{X}}      %% Frame field operator
\newcommand{\framemomop}{\op{\Pi}}      %% Frame field momentum operator
\newcommand{\wfunction}{\sigma}    %%Condensate wavefunction
\newcommand{\itwiner}{\mathcal{I}} %%Intertwiner
\newcommand{\spin}{j}      %%Spin rep
\newcommand{\spinrepdim}{d(\spin)}  %%Dimension of spin rep
\newcommand{\wmatrix}{D}   %%Wigner matrix
\newcommand{\peakfunc}{\eta} %%Peaking function
\newcommand{\cpeakwidth}{\epsilon} %%Clock Peaking width
\newcommand{\cpeakphase}{\pi_{0}}  %%Clock Peaking phase
\newcommand{\peakvalue}{x}         %%Peaking value
\newcommand{\peakphase}{\pi}         %%Peaking value
\newcommand{\redwfunction}{\tilde{\wfunction}}         %%Reduced condensate wavefunction
\newcommand{\kinetic}{K}       %% Kinetic term   
\newcommand{\metric}{g}         %%Metric
\newcommand{\lmetric}{\eta}         %%Lorentz metric
\newcommand{\rpeakwidth}{\delta} %%Rods Peaking width
\newcommand{\rpeakphase}{\pi_{x}}  %%Rods Peaking phase
\newcommand{\mommatterfield}{\pi_{\matterfield}}  %%Variable conjugate to \matterfield in Fourier transform 
\newcommand{\Kexpansindex}{s}  %%Expansion index of K
\newcommand{\binomexpansindex}{\ell}  %%Binomial expansion index
\newcommand{\derexpansindextime}{m} %%Time derivative expansion index
\newcommand{\derexpansindexspace}{n} %%Space derivative expansion index
\newcommand{\radius}{\varrho} 
\newcommand{\pangle}{\theta}       %%Polar angle 
\newcommand{\oangle}{\varphi}       %%Other angle in spherical coordinates 
\newcommand{\radiusint}{R}        %% Integral over radius
\newcommand{\polarint}{\Theta}        %% Integral over polar angle
\newcommand{\angularint}{\Phi}        %% Integral over angular variable
\newcommand{\combinedpeakparrods}{z}        %% rods peaking parameter combination entering in the equations of motion
\newcommand{\combinedpeakparclock}{z_{0}}        %% clock peaking parameter combination entering in the equations of motion
\newcommand{\normalcoeff}{\mathcal{N}}         %% normalization coefficient peaking function 
\newcommand{\kinratio}{r^{(\lambda)}}         %% Ratio of kinetic coefficients
\newcommand{\firstdercoeff}{\gamma}         %% Coefficient of the first derivative term 
\newcommand{\laplaciancoeff}{\alpha}       %% Coefficient of the spatial derivative terms
\newcommand{\nondercoeff}{{}^{(\lambda)}\!E}               %% Coefficient of the non-derivative term
\DeclareMathOperator{\re}{Re}   
\DeclareMathOperator{\im}{Im}   
\newcommand{\realpartnonder}{{}^{(\lambda)}\!\eta}               %% Real part of \nondercoeff
\newcommand{\impartnonder}{{}^{(\lambda)}\!\beta}               %% Imaginary part of \nondercoeff
\newcommand{\rwfunctionmodulus}{\rho}               %% Reduced condensate wavefunction modulus
\newcommand{\rwfunctionphase}{\theta}               %% Reduced condensate wavefunction phase
\newcommand{\bkgmodulus}{\bar{\rwfunctionmodulus}}               %% Background reduced condensate wavefunction modulus
\newcommand{\bkgphase}{\bar{\rwfunctionphase}}               %% Background reduced condensate wavefunction phase
\newcommand{\pertmodulus}{\delta\rwfunctionmodulus}               %% Perturbed reduced condensate wavefunction modulus
\newcommand{\pertphase}{\delta\rwfunctionphase}             %% Perturbed reduced condensate wavefunction phase
\newcommand{\massparameter}{\mu}             %% Mass parameter appearing in the background perturbation equation
\newcommand{\gravconst}{c}             %% Linear coefficient for the massparameter
\newcommand{\peakvaluemom}{\tilde{\pi}_{\phi}}             %% Peaking value in \mommatterfield
\newcommand{\mompeakingfunc}{f}             %% Peaking function in \mommatterfield
\newcommand{\mompeakingwidth}{\epsilon_{\matterfield}}             %% Peaking function width in \mommatterfield
\newcommand{\matterfieldop}{\op{\Phi}}      %% Matter field operator
\newcommand{\mattermomop}{\framemomop_{\matterfield}}      %% Matter field momentum operator
\newcommand{\coordinate}{x}      %% Coordinates
\newcommand{\bkgmatter}{\bar{\matterfield}}      %% Background matter field
\newcommand{\pertmatter}{\delta \matterfield}      %% Perturbed matter field
\DeclareMathOperator{\sgn}{sgn}
\newcommand{\oppkinetic}{\tilde{\kinetic}_{\lambda}}
\newcommand{\clockmomclassic}{\pi_{0}^{(c)}}
\newcommand{\bkgmattermomclassic}{\bar{\pi}_{\phi}^{(c)}}
\newcommand{\superhor}{(k\to 0)}
\newcommand{\gftaction}{S_{\text{GFT}}}

\newcommand{\normal}{X}
\DeclareMathOperator{\hyperbolic}{H^3}
\newcommand{\gvariablessl}{G_I}
\newcommand{\itwinerbc}{B}
\newcommand{\genlabel}{\upsilon}
\newcommand{\sgenlabel}{\upsilon_{o}}
\DeclareMathOperator*{\SumInt}{%
\mathchoice%
  {\ooalign{$\displaystyle\sum$\cr\hidewidth$\displaystyle\int$\hidewidth\cr}}
  {\ooalign{\raisebox{.14\height}{\scalebox{.7}{$\textstyle\sum$}}\cr\hidewidth$\textstyle\int$\hidewidth\cr}}
  {\ooalign{\raisebox{.2\height}{\scalebox{.6}{$\scriptstyle\sum$}}\cr$\scriptstyle\int$\cr}}
  {\ooalign{\raisebox{.2\height}{\scalebox{.6}{$\scriptstyle\sum$}}\cr$\scriptstyle\int$\cr}}
}
\newcommand{\generalrelativity}{General Relativity\xspace}
\newcommand{\quantumgravity}{quantum gravity\xspace}
\newcommand{\universe}{universe\xspace}
\newcommand{\qg}{QG\xspace}
\newcommand{\gr}{GR\xspace}
\newcommand{\GFTs}{GFTs\xspace}
\newcommand{\GFT}{GFT\xspace}
\newcommand{\TGFT}{TGFT\xspace}
\newcommand{\cosmology}{cosmology\xspace}
\newcommand{\LQG}{LQG\xspace}

\begin{document}

\title{Effective dynamics of scalar cosmological perturbations from  quantum gravity} %Force line breaks with \\

\author[a,b,c]{Luca Marchetti}
\author[b]{Daniele Oriti}

\affiliation[a]{Università di Pisa,\\Lungarno Antonio Pacinotti 43, 56126 Pisa, Italy, EU}
\affiliation[b]{Arnold Sommerfeld Center for Theoretical Physics,\\ Ludwig-Maximilians-Universit\"at München \\ Theresienstrasse 37, 80333 M\"unchen, Germany, EU}
\affiliation[c]{Istituto Nazionale di Fisica Nucleare sez. Pisa,\\Largo Bruno Pontecorvo 3, 56127 Pisa, Italy, EU}

\emailAdd{luca.marchetti@phd.unipi.it}
\emailAdd{daniele.oriti@physik.lmu.de}
%\thanks{}

\date{\today}

\abstract{We derive an effective dynamics for scalar cosmological perturbations from quantum gravity, in the framework of group field theory condensate cosmology. The emergent spacetime picture is obtained from the mean-field hydrodynamic regime of the fundamental theory, and physical observables are defined using a relational strategy applied at the same level of approximation, in terms of suitable collective states of the GFT field. The dynamical equations we obtain for volume and matter perturbations lead to the same solutions as those of classical \gr in the long-wavelength, super-horizon limit, but differ in other regimes. These differences could be of phenomenological interest and make contact between fundamental quantum gravity models and cosmological observations, indicating new physics or limitations of the fundamental models or of the approximations leading to the effective cosmological dynamics.}

\maketitle
\flushbottom
\section{Introduction}\label{sec:introduction}

In the last decades, cosmological observations have reached a remarkable level of precision, and have been shown to be compatible with a rather simple picture of the universe: a flat, (almost) Friedmann-Robertson-Walker (FRW) solution of \generalrelativity (\gr) which includes a positive cosmological constant $\Lambda$ and a cold dark matter component ($\Lambda$CDM paradigm) \cite{Planck:2018vyg}. 

Despite the simplicity of the resulting picture, the success of modern \cosmology is due to the interplay between different physical ingredients, each highly non-trivial \cite{Maggiore:2018sht,gorbunov1,gorbunov2,dodelson2020modern, Baumann:2009ds}: (i) \gr, describing how the fluids modelling the matter content of the \universe interact with gravity, in particular providing a framework in which the anisotropies and inhomogeneities that we observe today in the Cosmic Microwave Background (CMB) radiation and the large scale structures can be traced back to primordial ones \cite{Seljak:1996is}; (ii)  some detailed microscopic physics, describing the interactions (or lack of thereof) among the cosmological fluids (e.g. through the relativistic Boltzmann equation \cite{Cercignani2002}, a combination of the first two points); (iii) finally, inflation \cite{guth,sato,LINDE1982389} (or some alternative model of the very early universe \cite{Brandenberger:2018wbg}), providing a natural argument for the assignment of the initial conditions and a physical mechanism for the generation of the primordial perturbations acting as seeds of cosmic structure formation  (see e.g.\ \cite{Riotto:2002yw} for a review). It is then clear that, in order to extract cosmology in all of its glory from \quantumgravity (\qg), the theory needs to answer questions of very different physical nature. 

Any successful candidate \qg theory should be able to reproduce these results, in some approximation, and possibly help to clarify the nature of the exotic forms of matter and energy (e.g.\ the aforementioned cosmological constant and cold dark matter) which, surprisingly enough, compose the vast majority of our \universe. These represent established (in their observational consequences) but still rather mysterious (in their fundamental physical nature) ingredients of modern cosmology. Moreover, \qg is also expected to shed light on the very early phase of the universe, associated to the cosmological singularity predicted by \gr , where the whole semi-classical framework on which modern cosmology is based is likely to break down, and that it is simply not accounted for in current models. 
Moreover, this is all the more important, since \cosmology could be one of the best observational testing grounds for fundamental theories of \qg.

To extract cosmology from fundamental \qg formalisms is, in fact, a difficult challenge. This is especially true in \qg approaches based on structures that are not immediately related to continuum fields and that are formulated in a manifestly background independent manner.

Even just reproducing the purely \gr part of cosmological models (item (i) above) is far from being a simple task, for the following reasons. First of all, it requires the determination of appropriate continuum \emph{and} classical limits of the theory, two limits that are in general distinguished and independent in \qg \cite{Oriti:2018tym}. The continuum limit, in particular, requires control over the collective quantum dynamics and appropriate coarse-graining of the fundamental, microscopic degrees of freedom \cite{Oriti:2018dsg}, both being highly non-trivial tasks as experience with any quantum many-body system shows. Second, due to the absence of any manifold or spacetime structure in background independent \qg (a feature inherited directly from the classical background independence of \gr), time evolution and spatial localization in \qg cannot be defined in terms of the usual manifold structures on which effective field theory relies, and can only be intended in a relational sense, i.e., they can only be defined with respect to physical degrees of freedom. This relational strategy, already subtle at the classical level \cite{rovelliobservables}, becomes more tricky at the quantum level (see e.g.\ \cite{Hoehn:2019owq} for a detailed discussion). For quantum gravity theories suggesting an emergent spacetime scenario, this is even more true \cite{Marchetti:2020umh}.

Classically, a relational framework can be implemented through the use of \emph{relational observables} \cite{rovelliobservables, Dittrich:2004cb,Dittrich:2005kc} (see \cite{Tambornino:2011vg} for a review), gauge invariant extensions of phase space functions (associated to some physical quantities) which encode the relative change with respect to other phase space functions (associated to other physical quantities). For instance, a common choice to describe relational time evolution consists in minimally coupling the gravitational theory with a massless scalar clock (models of type II in \cite{Giesel:2012rb}), which is in fact well behaved enough to allow, for instance, for a reduced loop quantization of the system \cite{Domagala:2010bm}. When inhomogeneities, arguably the most important quantities for precision \cosmology measurements, are included in the picture, spatial rods need also to be employed. Again, this is commonly achieved by introducing simple matter degrees of freedom to be used as reference fields; in particular, one usually chooses models with four pairs of scalar degrees of freedom and second class constraints (models of type I in the notation of \cite{Giesel:2012rb}). Once the second class constraints are solved, four degrees of freedom are eliminated and the remaining ones are used as relational frame \cite{Giesel:2012rb}. An example of such models is the Brown-Kucha\v{r} dust introduced in \cite{Brown:1994py,Bicak:1997bx,Kuchar:1995xn}. They were used in fact in \cite{Giesel:2007wi,Giesel:2007wk} in order to define a cosmological perturbation theory in terms of relational quantities.

The very definition of relational evolution and localization in an emergent \qg theory (whose fundamental degrees of freedom are only indirectly related to continuum and classical quantities), however, is complicated by the fact that the quantities that we would classically manipulate in order to recast their relative change as relational evolution and localization are simply absent in the fundamental description. They are expected to be only available in the continuum limit, after an appropriate coarse graining of the microscopic degrees of freedom \cite{Marchetti:2020umh}.

These challenges are obviously not only technical, but also conceptual. 

In order to overcome them, one will need both guidance from physical non-\qg experience and a flexible \qg formalism. Group Field Theories (\GFTs) \cite{Krajewski:2012aw,Oriti:2011jm} offer both, and are thus promising candidate \qg frameworks where the possibility of extracting continuum cosmological physics from the fundamental theory is actually very concrete. They are quantum and statistical field theories defined on a group manifold (not interpreted as a \virgolette{spacetime}  manifold), typically characterized by non-local and combinatorial interactions. In this respect, they are generalizations of matrix models \cite{DiFrancesco:1993cyw,David:1992jw}, and as examples (together with, e.g., random tensor models \cite{Gurau:2011xp,Gurau:2016cjo,guraubook}) of models defined within a broader Tensorial Group Field Theory (\TGFT) formalism, i.e.\ tensorial field theory models which share the same non-local combinatorial pattern of interactions. More precisely, \GFTs are \TGFT models characterized by tensors whose data can be given a \virgolette{quantum geometric} interpretation. As a result of this structure, \GFTs offer interesting connections to other \qg approaches, like LQG \cite{Rovelli:2004tv,Thiemann:2007pyv,Ashtekar:2004eh}, spin foam models \cite{Perez:2003vx,Perez:2012wv,Finocchiaro:2018hks}, simplicial gravity models \cite{Finocchiaro:2018hks,Reisenberger:1997sk,Freidel:1998pt,Baratin:2011hp} and dynamical triangulations \cite{Ambjorn:2001cv,gorlich2013,Ambjorn2014,Loll:2019rdj}.

Because of their field theoretic nature, \GFTs offer tools and techniques that may prove helpful to tackle the above challenges. For instance, renormalization group techniques can be employed to study the continuum limit of the theory and the possible presence of phase transitions \cite{Carrozza:2016vsq,Finocchiaro:2020fhl, Pithis:2020kio,Pithis:2020sxm}. Alternatively, one could also employ a mean-field approach \cite{Oriti:2016qtz} to effectively describe the macroscopic dynamics (and also the critical behavior \cite{Pithis:2018eaq,Marchetti:2020xvf}) of the microscopic quantum gravitational many-body system. This perspective, which we will also adopt below, guides the extraction of cosmological physics from the hydrodynamics of \GFTs \cite{Oriti:2016acw}. In particular, this has been achieved in the recent literature by considering the mean-field dynamics of condensate states \cite{Gielen:2013naa,Gielen:2014ila,Gielen:2014uga,Oriti:2015qva,Gielen:2016dss,Oriti:2016qtz,Pithis:2016cxg,Pithis:2019tvp}, i.e.\ states characterized by the simplest possible collective behavior of the fundamental \GFT quanta (see however \cite{Gielen:2021vdd} for a more \virgolette{state agnostic} approach).

Due to their macroscopic properties, these states have also been used to implement an effective notion of relational evolution in \cite{Marchetti:2020umh} with respect to a massless scalar field clock. This effective notion of relationality, being defined only for emergent (and averaged) quantities, bypasses several technical and conceptual difficulties related to its definition for microscopic \qg degrees of freedom. Many intriguing results have been obtained from the effective relational \GFT condensate cosmology framework, by making use of an EPRL-like \GFT model (see \cite{Oriti:2016qtz} and Section \ref{sec:gftmodels} for more details). In particular, two regimes of the resulting emergent relational dynamics are worth mentioning \cite{Oriti:2016qtz,Marchetti:2020umh,Marchetti:2020qsq}. First, a continuum classical regime, characterized by a large number of \GFT quanta, which matches the flat Friedmann cosmological dynamics. Second, a bouncing regime, characterized by a possible (depending on the impact of quantum fluctuations and on initial conditions) averaged resolution of the cosmological singularity into a quantum bounce. Moreover, phenomenological studies on the \GFT interactions connected them to geometric inflation \cite{deCesare:2016rsf} and phantom dark energy \cite{Oriti:2021rvm}. These results have also been obtained recently using an extended Barrett-Crane (BC) model, which suggests that the emergent behavior of these theories may in fact be universal (at least at this level of approximation and for the few observables that have been considered so far) \cite{Jercher:2021bie}. 

Motivated by the success of these homogeneous and isotropic results, some pioneering works have tried making the first steps towards the study of small inhomogeneities \cite{Gielen:2017eco,Gielen:2018xph,Gerhardt:2018byq}. In particular, in \cite{Gielen:2017eco,Gielen:2018xph} it was explored the possibility for the production of primordial perturbations from quantum fluctuations of operators. However, the operators studied in \cite{Gielen:2017eco,Gielen:2018xph} did not yet have a solid relational interpretation. In \cite{Gerhardt:2018byq}, instead, the evolution of long wavelength perturbations was studied through the separate universe framework, already applied successfully also to Loop Quantum Gravity (\LQG) \cite{Wilson-Ewing:2015sfx}. Here, we aim to provide a generalization of the results obtained in \cite{Gerhardt:2018byq} also to smaller wavelengths and in terms of an effective localization of perturbations in terms of a proper relational matter frame consisting of four minimally coupled scalar fields. This kind of matter can be seen as the corresponding model of type II associated to the model of type I, and it was shown in \cite{Giesel:2016gxq} not to allow for a reduced loop quantization of the system. This, however, is not a restriction for our purposes, since we are only aiming for an effective relational description of the kinematic quantum gravity system, and not for a reduced phase space quantization.

The main objective of this work is to attempt to reproduce part (i) of the above list of ingredients making up modern cosmology, also for what concerns inhomogeneous cosmological perturbations (in both geometry and matter sectors), leaving e.g.\ the (equally important) task of reproducing part (iii) (which was instead the one considered by \cite{Gielen:2017eco,Gielen:2018xph}) to future works. More precisely, in Section \ref{sec:kinematics}, we will review the kinematics of the \GFT models we are interested in (i.e.\ EPRL-like and extended BC) and we will in particular introduce coherent states which are peaked in \virgolette{pre-matter} variables associated to the minimally coupled massless scalar fields we want to use as a physical frame. In Section \ref{sec:dynamics}, instead, we will specify the classical system we want to reproduce, whose matter content will be characterized by five minimally coupled massless scalar fields, four of which will make up the matter reference frame and will be assumed to have negligible contribution to the energy-momentum budget of the \universe. The remaining field, whose interplay with geometry dominates the resulting evolution of the \universe, will be assumed to include small inhomogeneities with respect to the matter frame. Moreover, in Section \ref{sec:dynamics}, we will also show how dynamical equations for the macroscopic quantities determining the condensate state can be obtained from a mean-field quantum \GFT dynamics. In Section \ref{sec:evophysicalquantities}, we will study how geometry and matter physical quantities evolve with respect to the matter fields frame, and we will discuss in particular the possibility of matching the results with \gr (in harmonic gauge) in an appropriate limit. The results will be discussed in Section \ref{sec:conclusions}, where we will also point out future research directions. Finally, in Appendix \ref{app:harmonicgauge} we have reviewed how a first order harmonic gauge can be imposed classically, while in Appendix \ref{sec:redwfunctiondynamics} we have reported the detailed computations leading to the results concerning dynamics of Section \ref{sec:dynamics}.
\section{\GFT effective relational cosmology: kinematics}\label{sec:kinematics}
In this section, we will review the basic notions of the \GFT formalism necessary for the extraction of effective relational cosmological dynamics. More precisely, in Section \ref{sec:gftmodels} we will first briefly review the definition of two models used in the literature for cosmological applications, i.e.,\ the EPRL-like and the extended Barret-Crane (BC) models. Then, we will discuss the Fock structure of these theories, in particular when minimally coupled massless scalar fields are included as additional degrees of freedom. 

We will then continue in Section \ref{sec:condensates} by introducing a certain class of states which can be associated, at least in some limit, to continuum geometries, and which can be in fact also used to define an effective notion of relationality, thus paving the way to the study of cosmological small relational inhomogeneities.
\subsection{GFT models and their Fock structure}\label{sec:gftmodels}
As mentioned in Section \ref{sec:introduction}, \GFTs are field theories describing a field $\gftfield:G^d\to \mathbb{C}$. The specific choice of the group manifold $G$, of the dimension $d$ and of the (combinatorial) action $\gftaction$, together with additional restrictions on the fields, characterize a given GFT model, as we will see explicitly below with two examples. These data are chosen so that the perturbative expansion of the partition function of the theory around the Fock vacuum can be matched with spinfoam or lattice gravity models. The amplitudes of such expansion, therefore, can be seen as discretized $\dimension$-dimensional spacetimes and geometries, with the group theoretic data characterizing the GFT being associated to discretized gravitational quantities. As a consequence of this construction, the boundary states of the theory, and thus the fundamental quanta of the theory, can be seen as $\dimension-1$-simplices. When $\dimension=4$, as we will consider from now on, these states can be seen as quantum tetrahedra whose geometric properties are encoded in the group-theoretic data. In this sense, in the GFT approach to QG the classical spacetime is expected to \virgolette{emerge} from the collective behavior of the fundamental \virgolette{pre-geometric quanta} of the theory. 
\paragraph{EPRL-like and extended BC. }
The extraction of continuum cosmological physics from \GFTs has been first obtained by considering an EPRL-like GFT model (see e.g.\ \cite{Oriti:2016qtz}). However, it has been recently shown \cite{Jercher:2021bie} that the vast majority of the results obtained within the EPRL-like model can be similarly obtained in an extended BC model. This would suggest that while the two models differ\footnote{For instance, the EPRL model, contrarily to the BC model, incorporates explicitly the Barbero-Immirzi parameter.} e.g.\ in the implementation of the simplicity constraint (see below), they still belong to the same continuum universality class, a feature already emphasized in \cite{Dittrich:2021kzs}.  

Here we will briefly review the kinematic structure of these models. With kinematic structure, we mean the kind of additional restrictions that are imposed on the GFT field in order to satisfy the so-called closure and the simplicity constraints. Geometrically, these constraints represent the fact that the bivectors associated to the faces of the fundamental tetrahedra sum to zero and are simple, respectively. These are nothing but the discrete counterparts of the imposition of gauge invariance and Plebanski geometricity condition in the continuum, respectively. How these conditions are imposed in a GFT characterizes the specific GFT model one is constructing. 
\begin{description}[font=\itshape]
\item[{Extended BC model}:] The extended BC model considered in \cite{Jercher:2021bie} is a Lorentzian version of the model defined in \cite{Baratin:2011tx}, which is in turn a generalization of the original BC model \cite{Barrett:1999qw,DePietri:1999bx,Perez:2000ec} which allows for a non-ambiguous (i.e.\ commuting) imposition of the closure and simplicity constraints. The group domain of the field is given by four copies of $\text{SL}(2,\mathbb{C})$, but it is extended to include a timelike normal $\normal\in \hyperbolic$, where $\hyperbolic=\text{SL}(2,\mathbb{C})/\text{SU}(2)$ is the $3$-hyperboloid, so $\gftfield(\gvariablessl)\to \gftfield(\gvariablessl;\normal)$, where $\gftfield(\gvariablessl;\normal)\equiv \gftfield(G_1,\dots, G_4;\normal)$, with each $\gvariablessl$ being an element of $\text{SL}(2,\mathbb{C})$.

Simplicity and closure are then defined with respect to the normal $\normal$ as follows:
\begin{subequations}\label{equations:closuresimplbc}
\begin{align}\label{eqn:simplicitybc}
    \gftfield(\gvariablessl;\normal)&=\gftfield(G_1g_1,\dots, G_4g_4;\normal)\,,\quad &\forall&\, g_I\in \text{SU}(2)_\normal\,,\\\label{eqn:closurebc}
    \gftfield(\gvariablessl;\normal)&=\gftfield(G_I h^{-1};h\cdot\normal)\,,\quad&\forall&\, h\in\text{SL}(2,\mathbb{C})\,,
\end{align}
\end{subequations}
where $\text{SU}(2)_\normal$ is the $\text{SU}(2)$ subgroup of $\text{SL}(2,\mathbb{C})$ stabilizing $\normal$. From the above expressions it is clear that the normal $\normal$ has the only purpose of defining consistently the above constraints. As such, as we will mention below, it is not a dynamical variable.
\item[{EPRL-like model}:] For cosmological applications, an EPRL-like GFT model has been implemented by following slightly different steps from the above extended BC model. Indeed, in the approach followed by \cite{Gielen:2013naa,Oriti:2016qtz}, one starts from a GFT defined on $\groupdomain=\text{SU}(2)$, with the details of the embedding of this $\text{SU}(2)$ subgroup inside $\text{SL}(2,\mathbb{C})$ (characterizing the appropriate imposition of the simplicity constraint of the model) being encoded in general in the kinetic and interaction terms of the action\footnote{In principle, the simplicity constraint can be imposed at the level of the kinetic term, of the interactions, or both \cite{Gielen:2013naa}. Each of these choices will in general result in different quantum theories. As we will see below, however, in this paper the precise details of the interaction and kinetic kernels will not be important so our results will encompass all the above choices.}. This still guarantees that the amplitudes of the perturbative expansion of the partition function of the model match those of the EPRL spinfoam model, but allows the use of kinematic structures which are easier to handle and, importantly, which offer a more direct geometric interpretation. 

Indeed, when the closure constraint is imposed similarly to equation \eqref{eqn:closurebc} (but without an explicit notion of normal):
\begin{equation}\label{eqn:closureeprl}
    \gftfield(\gvariables)=\gftfield(\gvariables h)\,,\qquad \forall h\in \text{SU}(2)\,,
\end{equation}
where $\gftfield(\gvariables)\equiv \gftfield(g_1,\dots, g_4)$ with each $g_I\in \text{SU}(2)$, the resulting boundary states and fundamental quanta of the theory can be seen as open spin-networks, i.e.\ nodes from which four links are emanating which are decorated with the equivalence class of geometrical data $[\{\gvariables \}]=\{\{\gvariables h\},h\in \text{SU}(2)\}$. This correspondence between the fundamental structures of the theory and spin-network allows for a straightforward connection with LQG, which may in particular prove helpful to gain insights for the extraction of continuum physics \cite{Oriti:2017ave}.
\end{description}
As mentioned above, we will from here on explain the basic ideas underlying the extraction of cosmology from \GFTs by working within an EPRL-like model for simplicity. However, we will emphasize similarities and differences between the two models where important, and we will resort to a unified notation where useful. 
\paragraph{Group representation basis.}
The interpretation we have provided above of the boundary states of the theory as spin-network states is even more clear when working in the spin representation. This can be done by expanding the field satisfying \eqref{eqn:closureeprl} on a basis of functions on $L^2(\groupdomain^4/\groupdomain)$, with $\groupdomain=\text{SU}(2)$. By denoting $\vspinrep=\{\spin_I,m_I,\iota\}$ the labels characterizing these basis functions, we have
\begin{equation}\label{eqn:gftfieldspinexpansion}
    \gftfield(\gvariables)=\sum_{\iota}\sum_{\spin_I}\sum_{m_I,n_I}\gftfield^{\spin_1,\dots,\spin_4;\iota}_{m_1,\dots,m_4}\left[\prod_{i=1}^4\sqrt{d(\spin_i)}D^{j_i}_{m_in_i}(\gvariables)\right]\itwiner^{\spin_1,\dots, \spin_4;\iota}_{n_1,\dots,n_4}\equiv \sum_{\vspinrep} \gftfield_{\vspinrep}\psi_{\vspinrep}(\gvariables)\,,
\end{equation}
where $\itwiner^{\spin_1,\dots, \spin_4;\iota}_{n_1,\dots,n_4}$ is an $\text{SU}(2)$ intertwiner obtained from the right-diagonal invariance of the GFT field. Exactly because of the choice $\groupdomain=\text{SU}(2)$, the boundary states are in fact clearly decorated with spin-network vertex data $\vspinrep$. More precisely, $\spin_I$ and $m_I$ are respectively spin and angular momentum projection associated to the open edges of a given vertex, while $\iota$ represents the intertwiner quantum number associated to the vertex itself. 

Of course, a similar decomposition can be performed for the GFT field operator of the extended BC model, the difference in this case being of course the field domain itself. The representation theory of $\groupdomain=\text{SL}(2,\mathbb{C})$ is clearly more involved than the one for $\text{SU}(2)$. However, for the purposes of this paper, we will only need some basic facts. Unitary irreducible representations of $\text{SL}(2,\mathbb{C})$ are labelled by $(\rho,\nu)$, with\footnote{Strictly speaking, this is only true for the \emph{principal series}, which we will restrict to here. For more details see \cite{Jercher:2021bie} and references therein. } $\rho\in\mathbb{R}$ and $\nu\in \mathbb{Z}/2$. The imposition of the simplicity constraint \eqref{eqn:simplicitybc} then forces $\nu=0$ \cite{Jercher:2021bie}. As a result, once integrating away the normal, one finds \cite{Jercher:2021bie}
\begin{equation}\label{eqn:integratedbcexpansion}
    \gftfield(\gvariablessl)\equiv \int_{\hyperbolic}\diff \normal \gftfield(\gvariablessl;\normal)=\int\diff\rho_I\sum_{\spin_I, l_I}\sum_{m_I,n_I}\gftfield^{\rho_I}_{\spin_I m_I}\left[\prod_{i=1}^4(4\rho_i^2)D^{(\rho_i,0)}_{\spin_i m_i l_in_i}(\gvariables)\right]\itwinerbc^{\rho_I}_{l_I n_I}\,,
\end{equation}
where $\diff\rho_I\equiv \prod_{i=1}^4\diff\rho_i$, $D^{(\rho_i,0)}_{\spin_i m_i l_in_i}(\gvariables)$ are representation matrices, with $\spin_i$ and $l_i$ being positive half-integers and $m_i\in \{-\spin_i,\dots, \spin_i\}$, $n_i\in \{-l_i,\dots,l_i\}$. Finally,  $\itwinerbc^{\rho_I}_{l_I n_I}$ is the Barret-Crane intertwiner,
\begin{equation}
    \itwinerbc^{\rho_I}_{l_I n_I}\equiv \int_{\hyperbolic}\diff X \prod_{i=1}^4D^{(\rho_i,0)}_{\spin_i m_i00}(X)\,.
\end{equation}
Notice that, being the intertwiner space one-dimensional, there is no intertwiner label $\iota$ in this case, contrarily to what we have seen in equation \eqref{eqn:gftfieldspinexpansion}. Despite this difference, equations \eqref{eqn:gftfieldspinexpansion} and \eqref{eqn:integratedbcexpansion} share many obvious similarities.
\paragraph{Fock structure.}
\GFTs can naturally be formulated in the language of second quantization. To this purpose, one defines the field operators $\gftfieldop$, $\gftfieldop^\dagger$ satisfying the commutation relations:
\begin{subequations}
\begin{align}\label{eqn:basiccommutator}
[\gftfieldop(\gvariables),\gftfieldop^\dagger(\gvariables')]&=\mathbb{I}_{\groupdomain}(\gvariables,\gvariables')\,,\\ [\gftfieldop(\gvariables),\gftfieldop(\gvariables')]&=[\gftfieldop^\dagger(\gvariables),\gftfieldop^\dagger(\gvariables')]=0\,,
\end{align}
\end{subequations}
where $\mathbb{I}_\groupdomain(\gvariables,\gvariables')$ is a Dirac delta distribution on the space $\groupdomain^4/\groupdomain$, with $\groupdomain=\text{SU}(2)$. For the extended BC model, instead, the commutation relations would read 
\begin{subequations}
\begin{align}\label{eqn:basiccommutatorbc}
[\gftfieldop(\gvariablessl;\normal),\gftfieldop^\dagger(\gvariablessl';\normal')]&=\mathbb{I}_{\text{BC}}(\gvariablessl;\normal,\gvariablessl';\normal')\,,\\ [\gftfieldop(\gvariablessl,\normal),\gftfieldop(\gvariablessl',\normal')]&=[\gftfieldop^\dagger(\gvariablessl,\normal),\gftfieldop^\dagger(\gvariablessl',\normal')]=0\,,
\end{align}
\end{subequations}
where, similarly as before, $\mathbb{I}_{\text{BC}}(\gvariablessl;\normal,\gvariablessl';\normal')$ is the identity on the space $L^2(\text{SL}(2,\mathbb{C})^4\times \hyperbolic)$ which preserves the symmetries \eqref{equations:closuresimplbc}. Upon quantization, the modes $\gftfield_{\vspinrep}$ of the decomposition \eqref{eqn:gftfieldspinexpansion} are quantized and become creation and annihilation operators $\gftfieldop_{\vspinrep}$ and $\gftfieldop^\dagger_{\vspinrep}$ for spin-network vertices, which satisfy
\begin{equation}
[\gftfieldop_{\vspinrep},\gftfieldop^\dagger_{\vspinrep'}]=\delta_{\vspinrep,\vspinrep'}\,,\qquad [\gftfieldop_{\vspinrep},\gftfieldop_{\vspinrep'}]=[\gftfieldop^\dagger_{\vspinrep},\gftfieldop^\dagger_{\vspinrep'}]=0\,.
\end{equation}
The Fock space is then constructed as usual from the repeated action of the creation operator on the vacuum state $\ket{0}$ annihilated by all $\gftfieldop_{\vspinrep}$s, whose $n$-particle states satisfy
\begin{align*}
 \gftfieldop_{\vspinrep}\ket{n_{\vspinrep}}&=\sqrt{n_{\vspinrep}}\ket{n_{\vspinrep}-1}\,,\\
 \gftfieldop_{\vspinrep}\ket{n_{\vspinrep}}&=\sqrt{n_{\vspinrep}+1}\ket{n_{\vspinrep}+1}\,.
 \end{align*}
It is possible to show that the Fock space constructed in this way shares many similarities with the kinematical Hilbert space of LQG \cite{Oriti:2013aqa}, since it encodes very similar degrees of freedom.

As usual in the second quantized approach, one can construct quantum observables out of the quantized field operators which act on states of the Fock space. The simplest example of such operators is the \emph{number operator}\footnote{For the extended BC case, as a consequence of the fact that the GFT field operator depends on the normal $\normal$, which is however non-dynamical, equation \eqref{eqn:numberoperator} becomes
\begin{equation*}
\op{N}\equiv \int_{\hyperbolic}\diff X\int\intmeasure{\gvariablessl}\gftfieldop^\dagger(\gvariablessl;\normal)\varphi(\gvariablessl;\normal)\mathperiod
\end{equation*}
}
\begin{equation}\label{eqn:numberoperator}
\op{N}\equiv \int\intmeasure{\gvariables}\gftfieldop^\dagger(\gvariables)\varphi(\gvariables)\,,
\end{equation}
whose eigenvalues characterize different sectors of the GFT Fock space, since they count the number of quanta present in a given state. A general second quantized operator then reads
\begin{equation}\label{eqn:secondquantizationoperator}
\op{O}_{n+m}\equiv \int (\diff \gvariables)^m(\diff h_{I})^n\,O_{m+n}(\gvariables^1,\dots, \gvariables^m,h_{I}^1,\dots, h_{I}^n)\prod_{i=1}^m\gftfieldop^\dagger(\gvariables^i)\prod_{j=1}^n\gftfieldop(h_{I}^j)\,,
\end{equation}
where the matrix elements $O_{m+n}$ can be obtained either from quantum simplicial considerations, or, in the case $\groupdomain=\text{SU}(2)$ that we are considering here, from the LQG matrix elements between spin-network vertex states. This construction is of course independent on the specific representation of the Hilbert space one chooses to work with. For instance, a generic one-body operator characterized by a creation and an annihilation operator can be written as
\begin{equation}\label{eqn:twobodyspin}
\op{O}_2=\sum_{\vspinrep\vspinrep'}O_2(\vspinrep,\vspinrep')\gftfieldop^\dagger_{\vspinrep}\gftfieldop_{\vspinrep'}\,.
\end{equation}
One-body operators, typically associated to macroscopic observables, are clearly of great interest for the extraction of emergent continuum physics, and will thus be those that we will focus on from here on.
\paragraph*{Coupling massless scalar fields.}
As discussed in Section \ref{sec:introduction}, in this paper we aim to describe relational cosmological inhomogeneities. To this purpose, it is necessary to identify a set of relational rods and clock to use as physical frame. A very simple choice of such physical frame consists in four minimally coupled free massless scalar fields\footnote{Indeed, as we will see in Section \ref{sec:dynamics}, this choice remarkably simplifies the quantum dynamics.}. Here, therefore, we introduce \virgolette{pre-matter} data alongside the purely geometric ones discussed above, which by construction can be associated to $n$ scalar fields. Indeed, the inclusion of such pre-matter degrees of freedom to the GFT is performed in such a way that the perturbative expansion of the GFT partition function matches the discrete path-integral of the simplicial gravity model minimally coupled with the massless scalar fields one wants to reproduce
 (see \cite{Li:2017uao} for more details). This clearly changes the precise form of the GFT action, which now has to take into account the discretized matter-geometry coupling, but it also impacts the kinematics of the GFT model. Indeed, consider as an example the case in which we want to include one single scalar field. Since the scalar field is \virgolette{discretized} on the simplicial structures corresponding to the GFT boundary states, the very same GFT field domain has to be enlarged in order to account for the additional matter data $\framefield\in\mathbb{R}$:   
\begin{equation}\label{eqn:includingonescalarfield}
\gftfieldop(\gvariables)\quad\longrightarrow\quad\gftfieldop(\gvariables,\framefield)\,, \qquad \mathcal{H}_1=L^2(\text{SU}(2)^4/\text{SU}(2))\to L^2(\text{SU}(2)^4/\text{SU}(2)\times \mathbb{R})\,,
\end{equation}
where $\mathcal{H}_1$ is the one-particle Hilbert space of the model. Notice that both these modifications to the dynamics and kinematics of the model happen regardless of the precise GFT model (e.g. EPRL-like or extended BC model). In particular, the arguments presented here in the exemplifying case of the EPRL-like model apply identically to the extended BC one.

Clearly, if one wants to introduce more (say $n$) than one minimally coupled massless scalar field, the group field operator becomes $\gftfieldop(\gvariables,\framefield^a)\equiv \gftfieldop(\gvariables,\framefield^1,\dots,\framefield^n)$, with $a=1,\dots,n$. Of course, the commutation relation in \eqref{eqn:basiccommutator} has to be changed consistently, so that 
\begin{equation}
    \left[\gftfieldop(\gvariables,\framefield^a),\gftfieldop^\dagger\left(h_I,(\framefield')^a\right)\right]=\mathbb{I}_G(\gvariables,h_I)\delta^{(n)}\left(\framefield^a-(\framefield')^a\right)\,.
\end{equation}
Importantly, this change on the kinematic structure of the Fock space is reflected also in the second quantized operators, which now involve integrals over all the possible values of $\framefield^a\in\mathbb{R}^n$. For instance, the number operator reads
\begin{subequations}
\begin{equation}
    \op{N}=\int\diff^n\framefield\int\intmeasure[]{\gvariables}\gftfieldop^\dagger(\gvariables,\framefield^a)\gftfieldop(\gvariables,\framefield^a)\,.
\end{equation}
A crucial quantity for describing cosmological geometries is the volume operator
\begin{equation}\label{eqn:volumeoperator}
    \op{V}=\int\diff^n\framefield\int\diff \gvariables\intmeasure[]{\gvariables'}\gftfieldop^\dagger(\gvariables,\framefield^a)V(\gvariables,\gvariables')\gftfieldop(\gvariables',\framefield^a)\,,
\end{equation}
whose matrix elements $V(\gvariables,\gvariables')$ are defined from those of the first quantized volume operator in the group representation\footnote{Such an operator is diagonal in the spin representation, with eigenvalues $\sim \spin^{3/2}$ for the EPRL-like model we are considering here and $\sim \rho^{3/2}$ for the extended BC model.}. 

The presence of \virgolette{pre-matter} data allows for the construction of a set of observables naturally related to them, through polynomials and derivatives with respect to $\chi^a$ for each $a=1,\dots, n$. In particular, the two fundamental, self-adjoint ones that can be obtained in this way are the \virgolette{scalar
field operator} and the \virgolette{momentum operator} \cite{Oriti:2016qtz}:
\begin{align}
\label{eqn:scalarfieldoperator}
\framefieldop^b&\equiv \int\diff^n\framefield\int\intmeasure[]{\gvariables}\framefield^b\gftfieldop^\dagger(\gvariables,\framefield^a)\gftfieldop(\gvariables,\framefield^a)\,,\\
\label{eqn:momentumoperator}
\framemomop_b&=\frac{1}{i}\int\diff^n\framefield\int\intmeasure[]{\gvariables}\left[\gftfieldop^\dagger(\gvariables,\framefield^a)\left(\frac{\partial}{\partial\framefield^b}\gftfieldop(\gvariables,\framefield^a)\right)\right],
\end{align}
\end{subequations}
whose expectation values on appropriate semi-classical and continuum states should be associated to the scalar field itself and possibly its momentum, which are at the core of a relational definition of dynamics and evolution \cite{Marchetti:2020umh}, as we will briefly review below.

\subsection{Continuum geometries, effective relationality and GFT condensates}\label{sec:condensates}
In order to describe the relational evolution of cosmological small inhomogeneities, one necessary step is to identify a class of quantum states which admit some \virgolette{proto-geometric} interpretation in terms of approximate continuum geometries. This allows to define an effective notion of relational evolution, whose general definition in a \virgolette{pre-geometric} sector of an emergent quantum gravity theory (such as a GFT) is instead technically and conceptually very complicated \cite{Marchetti:2020umh}, as we have discussed in Section \ref{sec:introduction}. Such \virgolette{proto-geometric} states are expected to be the result of some form of coarse-graining over the fundamental, microscopic degrees of freedom, and thus to show some form of collective behavior. In a sense, they are associated to a hydrodynamic description of the underlying quantum gravity model. The simplest form of such collective behavior is shown by \emph{coherent} (or, more commonly, \emph{condensate}) states, where each fundamental quantum is associated to the same condensate wavefunction:\label{asspage:KS1a}
\begin{subequations}
\begin{align}\label{eqn:coherentstates}
    \ket{\wfunction}&=\normalcoeff_{\wfunction}\exp\left[\int \diff^n \framefield\int\intmeasure[]{\gvariables}\wfunction(\gvariables,\framefield^a)\gftfieldop^\dagger(\gvariables,\framefield^a)\right]\ket{0}\\
    &=\normalcoeff_\wfunction\exp\left[\int \diff^n\framefield\sum_{\vspinrep}\wfunction_{\vspinrep}(\framefield^a)\gftfieldop_{\vspinrep}^\dagger(\framefield^a)\right]\ket{0}\,,\nonumber
\end{align}
\end{subequations}
where
\begin{subequations}
\begin{align}
\normalcoeff_{\wfunction}&\equiv e^{-\Vert \wfunction\Vert^2/2},\\\
\Vert\wfunction\Vert^2&=\int \diff^n \framefield\int\intmeasure[]{\gvariables}\vert\wfunction(\gvariables,\framefield^a)\vert^2\equiv \braket{ \op{N}}_{\wfunction}\,.
\end{align}
\end{subequations}
By definition, such coherent states are eigenstates of the field operator:
\begin{equation}\label{eqn:eigenstateannihilation}
    \gftfieldop(\gvariables,\framefield^a)\ket{\wfunction}=\wfunction(\gvariables,\framefield^a)\ket{\wfunction}\,,\qquad \gftfieldop_{\vspinrep}(\framefield^a)\ket{\wfunction}=\wfunction_{\vspinrep}(\framefield^a)\ket{\wfunction}\,.
\end{equation}
States of the form \eqref{eqn:coherentstates} have been used in the past literature to show the intriguing results about the extraction of homogeneous and isotropic cosmological physics from \GFTs mentioned in Section \ref{sec:introduction}. Moreover, they allow for a simple implementation of an effective description of relational quantities, as we explain below.
\paragraph{Symmetries of the condensate wavefunction.}
Before discussing how an effective relational framework can be implemented, let us mention some important symmetry assumptions that are often made on the condensate wavefunction. Let us also emphasize that the imposition of symmetry properties of the condensate wavefunction is conceptually different from a symmetry reduction procedure. Indeed, the first is a condition on a collective macroscopic quantity, while the latter acts on the fundamental microscopic degrees of freedom (though technically in the case of a coherent state like the one in \eqref{eqn:coherentstates} the collective wavefunction is also the wavefunction of each microscopic tetrahedron).

A first important symmetry that is imposed on the condensate wavefunction is a diagonal left-invariance:
\begin{equation}
    \wfunction(\gvariables,\framefield^a)=\wfunction(h\gvariables,\framefield^a)\,,\qquad \forall h\in\text{SU}(2)\,.
\end{equation}
This condition can be seen as an average over the relative embedding
of the tetrahedron in $\mathfrak{su}(2)$ \cite{Oriti:2016qtz}. As a consequence of this imposition, the domain of the condensate wavefunction is isomorphic to the space of all the spatial metrics at a point, or, equivalently, to minisuperspace \cite{Gielen:2014ila}. This very same result holds also in the case of the extended BC model, with a similar averaging procedure (now over all configurations involving a preferred hypersurface normal and thus only for the integrated condensate wavefunction with respect to the normal $\normal$) \cite{Jercher:2021bie}.  

An additional assumption that is often imposed on the condensate wavefunction is its isotropy \cite{Oriti:2016qtz} (assumption \ref{ass:ks2})\label{asspage:ks2a}. This drastically simplifies the continuum dynamics, since the condensate wavefunction effectively turns out to depend on only one spin label $\spin$:
\begin{equation}\label{eqn:isotropycond}
    \wfunction(\gvariables,\framefield^a)=\sum_{\spin=0}^\infty\wfunction_j(\framefield^a){\itwiner^*}^{\spin\spin\spin\spin,\iota_+}_{m_1m_2m_3m_4}\itwiner^{\spin\spin\spin\spin,\iota_+}_{n_1n_2n_3n_4}\sqrt{\spinrepdim^4}\prod_{i=1}^4\wmatrix^\spin_{m_in_i}(\gvariables)\,,
\end{equation}
where $\spinrepdim=2\spin+1$ and $\iota_+$ is the largest eigenvalue of the volume operator compatible with $\spin$. Therefore, the condensate wavefunction in spin representation reads 
\begin{equation}\label{eqn:sigmachi}
    \wfunction_{\vspinrep}(\framefield^a)\equiv \wfunction_{\{\spin,\vec{m},\iota_+\}}(\framefield^a)=\wfunction_{\spin}(\framefield^a)\overline{\mathcal{I}}^{\spin\spin\spin\spin,\iota_+}_{m_1m_2m_3m_4}\,.
\end{equation}
Importantly, a similar result holds also for the extended BC model, with the dynamical part of the condensate wavefunction $\wfunction_{\rho}(\framefield^a)$, effectively depending on the continuous representation label $\rho$ \cite{Jercher:2021bie}. As a consequence, it is useful to define a label $\genlabel$ which can be identified with $\rho$ or $\spin$ depending on the specific model chosen (extended BC or EPRL-like, respectively). 

For instance, in terms of this new label, we can write the expectation value of the number and the volume operators on an isotropic coherent state as
\begin{equation}\label{eqn:volumenumberspin}
\braket{\op{N}}_{\wfunction}=\SumInt_{\genlabel} \vert \wfunction_{\genlabel}(\framefield^a)\vert^2\,,\qquad
    \braket{\op{V}}_{\wfunction}=\SumInt_{\genlabel} V_{\genlabel}\vert \wfunction_{\genlabel}(\framefield^a)\vert^2\,,
\end{equation}
with the $\SumInt$ symbol indicating that, depending on whether $\genlabel=\rho$ or $\genlabel=\spin$, the right-hand-sides of the above equations will involve an integral or a sum, respectively.
\paragraph{Effective relationality in GFT: CPSs.}\label{asspage:ks3a}
Let us now discuss a way to implement an effective relational description of physical quantities in the GFT formalism. As we have mentioned above, in \cite{Marchetti:2020umh} an effective framework for the relational evolution of geometric observables with respect to a scalar field clock was constructed by making use of Coherent Peaked States (CPSs). As the name suggests, these are coherent states of the form \eqref{eqn:coherentstates} whose wavefunction has however strong peaking properties on the scalar field variables. For instance, for a single scalar field clock, we would have 
\begin{equation}\label{eqn:wavefunctioncps}
 \wfunction_{\cpeakwidth}(\gvariables,\framefield^0)\equiv \peakfunc_{\cpeakwidth}(\gvariables;\framefield^0-\peakvalue^0,\cpeakphase)\redwfunction(\gvariables,\framefield^0)\,,
 \end{equation} 
where the peaking properties around $\peakvalue^{0}$ are encoded in the \emph{peaking function} $\peakfunc_{\cpeakwidth}$ with a typical width given by $\cpeakwidth$. Of course, in order for the peaking properties to be effective, one wants $\cpeakwidth$ to be very small, $\cpeakwidth\ll 1$. However, one cannot just take $\cpeakwidth\to 0$, because as a consequence of the Heisenberg uncertainty principle, the fluctuations of the operator $\framemomop^0$ defined in equation \eqref{eqn:momentumoperator} would become arbitrarily large, which is certainly not ideal if one wants to interpret the scalar field as a classical clock, at least in some appropriate limit. In order to guarantee the existence of such classical clock regime, in \cite{Marchetti:2020umh,Marchetti:2020qsq} the condensate wavefunction \eqref{eqn:wavefunctioncps} was also assumed to be dependent on the parameter $\cpeakphase$, satisfying $\cpeakwidth\cpeakphase^2\gg 1 $. As a concrete example of the above peaking function, one can consider a Gaussian \cite{Marchetti:2020umh,Marchetti:2020qsq}:
\begin{equation}\label{eqn:peakingfunction}
\peakfunc_{\cpeakwidth}(\framefield^0-\peakvalue^{0},\cpeakphase)\equiv \normalcoeff_{\cpeakwidth}\exp\left[-\frac{(\framefield^0-\peakvalue^{0})^2}{2\cpeakwidth}\right]\exp[i\cpeakphase(\framefield^0-\peakvalue^{0})]\,,
\end{equation}
with a normalization constant $\normalcoeff_{\cpeakwidth}$ and where, as a first ansatz, it was assumed that the peaking function is independent on the group variables $\gvariables$. Therefore, the \emph{reduced wavefunction} $\redwfunction$ (which is assumed not to spoil the peaking properties of $\peakfunc_{\cpeakwidth}$) encodes all the geometric properties of the state, and in particular shows the symmetries discussed above. Its specific form will be determined by dynamical considerations in Section \ref{sec:dynamics}. By construction, these states satisfy, in the limit of small $\cpeakwidth$ (see \cite{Marchetti:2020umh,Marchetti:2020qsq} for more details):
\begin{equation}
    \braket{\hat{\framefield}^0}_{\sigma}\equiv \frac{\braket{\framefieldop^0}_{\sigma}}{\braket{\hat{N}}_{\sigma}}\simeq \peakvalue^0\,,
\end{equation}
where the expectation value in the above equation is computed with respect to states with condensate wavefunction given by \eqref{eqn:wavefunctioncps}. Notice that we are defining a scalar field operator here through the intensive version of the second quantized operator $\framefieldop^0$ in equation \eqref{eqn:scalarfieldoperator}. As a consequence of the above equation, change with respect to $\peakvalue^0$ can be associated to evolution with respect to the \emph{averaged} clock. In this sense, the CPSs realize a notion of effective relational evolution of averaged geometric quantities with respect to the physical clock ($\peakvalue^0)$, with the effective approach becoming increasingly accurate the smaller one takes $\cpeakwidth$ and the larger $\braket{\op{N}}_{\sigma}$ grows. Indeed, it has been shown in \cite{Marchetti:2020qsq} that this latter condition, typically associated to an emergent regime and dynamically related to a very large value of the relational clock $\vert \peakvalue^0\vert$, is what ultimately allows for a suppression of fluctuations of both geometric and clock variables. In this large $\braket{\op{N}}_{\sigma}$ regime, the evolution of expectation values of geometric quantities with respect to $\peakvalue^0$ can thus be interpreted as classical relational dynamics\footnote{\label{footnote:fluctuations}On the other hand, when $\braket{\op{N}}_{\sigma}$ cannot be taken to be large, quantum fluctuations on both geometric and clock variables may become large, thus suggesting that the averaged evolution at the core of the effective relational approach may not be reliable anymore \cite{Marchetti:2020qsq}.}.

The physical interpretation of the CPSs is then clear: they assign a distribution of spatial geometries for each value of the peaking parameter $\peakvalue^0\in\mathbb{R}$, i.e., to each value the scalar field clock takes on average. As such, in the emergent limit of large number of particles (and $\cpeakwidth\ll 1$) one can see them as the quantum equivalent of leaves of a  $\framefield$-foliation of spacetime.

A similar construction can be performed if one is interested in describing also relational inhomogeneities of physical quantities. Assume that the spacetime dimension is $\dimension$, with $\dimension\le n$. Then, the condensate state with condensate wavefunction given by
\begin{equation}\label{eqn:patchstates}
    \wfunction_{\cpeakwidth^{\mu},\peakphase_\mu;\peakvalue^\mu}(\gvariables,\framefield^a)=\left[\prod_{\mu=0}^{\dimension-1}\peakfunc_{\cpeakwidth^\mu}(\framefield^\mu-\peakvalue^\mu,\peakphase_\mu)\right]\redwfunction(\gvariables,\framefield^a)\,,
\end{equation}
where the peaking function\footnote{Here we are assuming, as for the single scalar field case, that $\cpeakwidth^\mu\ll 1$ and $\cpeakwidth^\mu \peakphase_\mu^2\gg 1$ for each $\mu=0,\dots,\dimension-1$.} $\peakfunc_{\cpeakwidth^\mu}(\gvariables;\framefield^\mu-\peakvalue^\mu,\peakphase_\mu)$ can be taken to a Gaussian as in equation \eqref{eqn:peakingfunction} for each $\mu=0,\dots,\dimension-1$, would encode the distribution of spatial geometric data for each point $\peakvalue^\mu$ of the physical manifold coordinatized by the frame fields $\framefield^\mu$. By construction, the expectation value of the intrinsic version of the second quantized field operators $\framefieldop^\mu$ in equation \eqref{eqn:scalarfieldoperator} on the above states is approximately given by
\begin{equation}
    \braket{\hat{\framefield}^\mu}_{\sigma}\equiv \frac{\braket{\framefieldop^\mu}_{\sigma}}{\braket{\hat{N}}_{\sigma}}\simeq \peakvalue^\mu\,,
\end{equation}
thus characterizing the change with respect to $\peakvalue^\mu$ as physical. These will be the fundamental states that we will consider from now on. Before concluding this discussion, let us also emphasize that the implementation of relational evolution through the CPSs (and thus, also their physical interpretation) that we have reviewed here for an EPRL-like model, can be identically realized also for the extended BC model, with the simple substitution $\gvariables\to (\gvariablessl;\normal)$ in all the above equations \cite{Jercher:2021bie}.
\section{GFT effective relational cosmology:  dynamics}\label{sec:dynamics}
The main aim of this section is to obtain the dynamical equations which, once solved, determine the specific form of the reduced condensate wavefunction $\redwfunction$. The microscopic GFT action $\gftaction$ determining these equation is in turn obtained by comparison with an appropriate simplicial gravity model (see e.g.\ the discussion in Section \ref{sec:gftmodels}). Therefore, in Section \ref{sec:classicalsystem}, we will specify which kind of classical system we are interested in. Then, in Section \ref{sec:gftaverageddyna} we will obtain the dynamical equations determining the evolution of the reduced condensate wavefunction from the imposition of averaged GFT quantum equations of motion. Finally, in Section \ref{sec:perturbations}, we will define background and perturbed quantities, and we will consistently study the dynamical equations at zeroth and first order in the small perturbations.
\subsection{Classical system}\label{sec:classicalsystem}
The system we want to describe is classically composed by $\dimension+1$ massless scalar fields minimally coupled to gravity. We also assume that $\dimension$ of these fields, which we call $\framefield^\mu$, $\mu=0,\dots,\dimension-1$, give a negligible contribution to the total energy-momentum tensor of the system, while the contribution coming from the remaining scalar field, which we call $\matterfield$, is dominant (assumption \ref{ass:ds4})\label{asspage:ds4}. The $d$ scalar fields $\framefield^\mu$, therefore, can be thought as \virgolette{test fields} which we would naturally use to define a material reference system, for instance using harmonic coordinates $\coordinate^\mu$ (see Appendix \ref{app:harmonicgauge} for more details). The field $\matterfield$ is assumed to be almost homogeneous with respect to the material coordinate system (or, equivalently to the harmonic frame), meaning that $\matterfield=\bkgmatter+\pertmatter$, with $\bkgmatter=\bkgmatter(t)$, $t\equiv \coordinate^0$, being the homogeneous component of the field.

\paragraph{Matter action and symmetries.} 
At the classical level, therefore, we assume a matter action of the form
\begin{subequations}
\begin{align}
    S_m[\framefield^\mu,\matterfield]&=-\frac{1}{2}\int\diff^4x\sqrt{-g}\metric^{ab}\partial_{a}\framefield^0\partial_b\framefield^0+\frac{\lambda}{2}\sum_{i=1}^d\int\diff^4x\sqrt{-g}\metric^{ab}\partial_{a}\framefield^i\partial_b\framefield^i\nonumber\\
    &\quad-\frac{\alpha_\phi}{2}\int\diff^4x\sqrt{-g}\metric^{ab}\partial_{a}\matterfield\partial_b\matterfield\,,\label{eqn:matteraction}\\
    &=\frac{1}{2}\int\diff^4x\sqrt{-g}M^{(\lambda)}_{\mu\nu}\metric^{ab}\partial_{a}\framefield^\mu\partial_b\framefield^\nu-\frac{\alpha_\phi}{2}\int\diff^4x\sqrt{-g}\metric^{ab}\partial_{a}\matterfield\partial_b\matterfield\,,
    \end{align}
\end{subequations}
where $\alpha_\phi\gg 1$, and $\lambda=\pm 1$, so that $M^{(+1)}_{\mu\nu}=\lmetric_{\mu\nu}$, while $M^{(-1)}_{\mu\nu}=-\delta_{\mu\nu}$. The choice $\lambda = -1$ corresponds to the natural matter coupling of four free, massless and minimally coupled scalar fields in classical gravity, in which all of them are treated on identical footing. Moreover, only when $\lambda=-1$ one is guaranteed to have the appropriate sign for the energy (density) of the four fields. On the other hand, when $\lambda=+1$, the second term in the action has an opposite sign with respect to the first and the third one. As we will see below, the \GFT action is constructed so that it respects the symmetries of the classical matter action \cite{Oriti:2016qtz,Marchetti:2020umh, Gielen:2017eco}, and in particular it is symmetric under a rotation of the four scalar fields, with signature of the rotation depending the value of $\lambda$. Given this, one would expect the choice $\lambda=+ 1$ to the appropriate one for the interpretation of the above four scalar fields as a relational frame. However, as we have argued above, such an interpretation is available only at an emergent level and in an effective sense; as a consequence, only the symmetry properties of macroscopic variables (entering in the definition the CPSs) actually determine the emergent signature of the effective frame, which may well be independent of the signature of the orthogonal transformation relating the fields in the action. In order to see this clearly below, and to keep things as general as possible, leaving it to the effective cosmological dynamics to fix some of the ambiguities, we will not restrict to a specific choice of $\lambda=\pm 1$, keeping any $\lambda$-dependence explicit.

As we have just mentioned, the symmetries of the above action play an important role in determining the form of the GFT action. These are (cfr. \cite{Gielen2020}):
\begin{description}[font=\itshape]
\item[{Translations}:] $\framefield^\mu\to \framefield^\mu+ k^\mu$ and $\matterfield\to\matterfield+k$, for each $\mu=0,\dots,\dimension-1$.
\item[{Reflections}:] $\framefield^\mu\to -\framefield^\mu$ and $\matterfield\to -\matterfield$, for each $\mu=0,\dots, \dimension-1$.
\item[{Lorentz transformations/Euclidean rotations}] When $\lambda=+1$ (resp.\ $\lambda=-1$), transformations $R\in \text{SO}(1,3)$ (resp.\ \text{SO}(4)) acting as $\framefield^\mu\to \tensor{R}{^\mu_\nu}\framefield^\nu$ are a symmetry of the Lagrangian for each $\mu=0,\dots, \dimension-1$.
\end{description}
\subsection{GFT averaged dynamics}\label{sec:gftaverageddyna}
Analogously to what has been done in \cite{Oriti:2016qtz,Marchetti:2020qsq}, here we will only extract an effective mean-field dynamics from the full quantum equations of motion. In other words, we will only consider the imposition of the quantum equations of motion averaged on the states that we consider to be relevant for an effective relational description of the cosmological system (assumption \ref{ass:ds2})\label{asspage:ds2}, which, in our case, would be coherent states  $\ket{\wfunction_{\cpeakwidth^\mu};\peakvalue^\mu,\peakphase_{\mu}}$ as in equation \eqref{eqn:coherentstates} whose condensate wavefunction is assumed to take the form \eqref{eqn:patchstates} (assumptions \ref{ass:ks1} and \ref{ass:ks3})\label{asspage:KS1b}: 
\begin{align}\label{eqn:simplestschwinger}
&\left\langle\frac{\delta \gftaction[\gftfieldop,\gftfieldop^\dagger]}{\delta\gftfieldop^\dagger(\gvariables,\peakvalue^\mu)}\right\rangle_{\wfunction_{\cpeakwidth^\mu};\peakvalue^\mu,\peakphase_{\mu}}\equiv\left\langle\wfunction_{\epsilon^\mu};\peakvalue^\mu,\peakphase_{\mu}\biggl\vert\frac{\delta \gftaction[\gftfieldop,\gftfieldop^\dagger]}{\delta\gftfieldop^\dagger(\gvariables,\peakvalue^\mu)}\biggr\vert\wfunction_{\cpeakwidth^\mu};\peakvalue^\mu,\peakphase_{\mu}\right\rangle=0\,,
\end{align}
Here, $\gftaction$ is the GFT action, whose specific form will be discussed below. While perfectly consistent with the effective and approximate nature of the relational framework discussed in the previous section, the imposition of only an averaged form of equations of motion is clearly a strong truncation of the microscopic quantum dynamics, which is expected to be justified in general only in the emergent regime of very large number of particles (see the disucssion in Section \ref{sec:condensates} and in footnote \ref{footnote:fluctuations}).

Moreover, for the purposes of this work, we will be interested in observables capturing only isotropic perturbations (e.g.\ the volume operator \eqref{eqn:volumeoperator}). For this reason, not only we will assume that the reduced wavefunction is isotropic, in the sense explained in Section \ref{sec:condensates} (so that the expectation value of the volume operator reduces to \eqref{eqn:volumenumberspin}), but we will also consider a condensate state whose peaking properties are isotropic as well (assumption \ref{ass:kc2})\label{asspage:kc2a}:
\begin{equation}\label{eqn:condensatewavefunction}
      \wfunction_{\cpeakwidth,\rpeakwidth,\cpeakphase,\rpeakphase;\peakvalue^\mu}(\gvariables,\framefield^\mu,\matterfield)=\peakfunc_{\cpeakwidth}(\framefield^0-\peakvalue^0;\cpeakphase)\peakfunc_{\rpeakwidth}(\vert \framefieldbf-\mathbf{\peakvalue}\vert;\rpeakphase)\redwfunction(\gvariables,\framefield^\mu,\matterfield)\,,
\end{equation}
where $\vert \framefieldbf-\mathbf{\peakvalue}\vert^2=\sum_{i=1}^d(\framefield^i-\peakvalue^i)^2$. For the moment we will also assume (assumption \ref{ass:kc1})\label{asspage:kc1a} that the parameter $\rpeakwidth$ is a complex quantity, $\mathbb{C}\ni \rpeakwidth=\rpeakwidth_r+i\rpeakwidth_i$, but with a positive real part, necessary for the peaking properties of the states, $\rpeakwidth_r>0$. As we will see below, allowing a complex width for the rods peaking function allows the perturbation equations to be dependent on a derivative kernel with emergent Lorentz signature.
\paragraph{GFT action.}
Having made these premises, we now specify the form of $\gftaction$. As explained in Section \ref{sec:gftmodels}, $\gftaction$ depends on the precise spinfoam (or simplicial gravity) model coupled with $d+1$ massless scalar fields one wants to reproduce. While the EPRL-like and extended BC models differ on their domain (respectively $\text{SU}(2)$ and $\text{SL}(2,\mathbb{C})\times \hyperbolic$) and on the precise way the simplicity constraint is imposed, thus resulting in (in principle) different kinetic and interaction kernels, they are both defined by an action including a quadratic kinetic term and a non-local interaction term $U+U^*$ (the star representing complex conjugation) of simplicial\footnote{These kind of interactions are called simplicial because they represent the gluing of $5$ different tetrahedra in order to form a $4$-simplex, the basic building block of a $4$-dimensional discretized manifold.} type characterized by $5$ powers of the field operator, $\gftaction=\kinetic+U+U^*$.

The resulting form of the action is however quite complicated to handle for most practical applications. For this reasons, one often makes some additional simplifying assumptions on $\gftaction$ \cite{Oriti:2016qtz,Marchetti:2020umh}:
\begin{itemize}
    \item First of all, one imposes that the field symmetries of the classical action are preserved at the quantum level, meaning that they are also symmetries of the GFT action $\gftaction$ (assumption \ref{ass:ds1})\label{asspage:ds1}. In the case considered here, the symmetries to be respected are those highlighted in the section above: invariance under Lorentz transformations/Euclidean rotations, shifts, and reflections. This greatly simplifies the form of the interaction and kinetic terms, which read, in the EPRL-like case\footnote{Similar expressions hold for the extended BC model, provided that one extends the domain of the GFT fields and kinetic interaction kernels as $\gvariables\to (\gvariablessl;\normal)$. Moreover, since the normal $\normal$ is non-dynamical, the interaction kernel does not depend on it. As a consequence, only the integrated field \eqref{eqn:integratedbcexpansion} becomes important at the level of interactions. The kinetic kernel instead depends on the normal in a localized way, imposing $\normal=\normal'$, with $\normal$ and $\normal'$ being the arguments of $\bar{\gftfield}$ and $\gftfield$ respectively. We refer to \cite{Jercher:2021bie} for more details on the action of the extended BC model.} \cite{Oriti:2016qtz,Marchetti:2020umh}
    \begin{align*}
    \kinetic&=\int\diff \gvariables\diff h_I\int \diff^d\framefield\diff^d\framefield'\diff\matterfield\diff\matterfield'\,\bar{\varphi}(\gvariables,\framefield)\mathcal{K}(\gvariables,h_I;(\framefield-\framefield')^2_\lambda,(\matterfield-\matterfield')^2)\varphi(h_I,(\framefield')^\mu,\matterfield')\,,\\
    U&=\int\diff^d\framefield\diff\matterfield\int\left(\prod_{a=1}^5\diff \gvariables^a\right)\mathcal{U}(\gvariables^1,\dots,\gvariables^5)\prod_{\ell=1}^5\varphi(\gvariables^\ell,\framefield^\mu,\matterfield)\,,
\end{align*}
where $(\framefield-\framefield')_\lambda^2\equiv \sgn(\lambda) M^{(\lambda)}_{\mu\nu}(\framefield-\framefield')^\mu(\framefield-\framefield')^\nu$ and where $\mathcal{K}$ and $\mathcal{U}$ are the respectively the aforementioned kinetic and interaction kernels encoding information about the EPRL-like model and, in particular, about the specific Lorentzian embedding of the theory.
\item The second simplifying assumption that is often made in cosmological applications is that one is interested in a \virgolette{mesoscopic regime} where interactions are in fact essentially negligible (assumption \ref{ass:ds3})\label{asspage:ds3}. Clearly, this can only be a transient regime, and one expects that, eventually, interactions do become important (see e.g.\ \cite{Pithis:2016wzf,Pithis:2016cxg, Oriti:2021rvm}, for some works which study the phenomenological implications of the inclusion of interactions).
\end{itemize}
\paragraph{Dynamical equations.}
Under both these assumptions, and performing a Fourier transform with respect to the variables $\matterfield$ and $\matterfield'$, one can see that the averaged quantum equations of motion reduce to
\begin{equation}\label{eqn:fundamentalequationscps}
\int\diff h_I\int\diff^d\framefield\, \kinetic(\gvariables,h_I;\framefield^2_\lambda,\mommatterfield)\peakfunc_\cpeakwidth(\framefield^0;\cpeakphase)\peakfunc_{\rpeakwidth}(\vert\framefieldbf\vert;\rpeakphase)\redwfunction(h_I,\framefield^0+\peakvalue^0,\framefieldbf+\mathbf{x},\mommatterfield)=0\,,
\end{equation}
where $\mommatterfield$ is the variable canonically conjugate to $\matterfield$ with respect to the Fourier transform. Expanding $\kinetic$ and $\redwfunction$ in power series around $\framefield^0=0$, $\framefieldbf=0$ \cite{Marchetti:2020umh}, and assuming that (i) $\vert \rpeakwidth\vert$ and $\cpeakwidth$ are small, but the quantities 
\begin{equation}
    \combinedpeakparclock \equiv \cpeakwidth\cpeakphase^2/2\,,\qquad \combinedpeakparrods\equiv \rpeakwidth\rpeakphase^2/2
\end{equation}
are large in absolute value (assumption \ref{ass:ks3})\label{asspage:ks3b} and (ii) reducing to isotropic configurations (assumption \ref{ass:ks2})\label{asspage:ks2b}, one  finds, at the lowest order in the small parameters $\vert \rpeakwidth\vert$ and $\cpeakwidth$ (see Appendix \ref{sec:redwfunctiondynamics} for a detailed derivation)\label{asspage:kc2b}:
\begin{equation}\label{eqn:redwfunctionevoprio}
\partial^2_0\redwfunction_{\spin}(x,\mommatterfield)-i\firstdercoeff\partial_{0}\redwfunction_{\spin}(x,\mommatterfield)-\nondercoeff_{\spin}^2(\mommatterfield)\redwfunction_{\spin}(x,\mommatterfield)+\laplaciancoeff^2\nabla^2\redwfunction_{\spin}(x,\mommatterfield)=0\,,
\end{equation} 
where $\spin$ is the isotropic spin label introduced in equation \eqref{eqn:sigmachi}, where we have dropped the superscript ${}^\mu$ for the argument of the reduced wavefunction $\redwfunction$, $\peakvalue\equiv \peakvalue^\mu$  and where $\partial^2_0$ and $\nabla^2\equiv \sum_{i}\partial^2_i$ represent derivatives with respect to rod and clocks values respectively. Finally, we have defined
\begin{equation*}
\firstdercoeff\equiv \frac{\sqrt{2\cpeakwidth}\combinedpeakparclock}{\cpeakwidth \combinedpeakparclock^2}\,,\qquad \nondercoeff_{\spin}^2\equiv\frac{1}{\cpeakwidth \combinedpeakparclock^2}-\kinratio_{\spin;2}(\mommatterfield)\left(1+3\lambda \laplaciancoeff^2\right)\,,\qquad \laplaciancoeff^2\equiv \frac{1}{3}\frac{\rpeakwidth \combinedpeakparrods^2}{\cpeakwidth \combinedpeakparclock^2}\,,\qquad \kinratio_{\Kexpansindex}\equiv \frac{\oppkinetic^{(\Kexpansindex)}}{\oppkinetic^{(0)}}\,.
\end{equation*}
Notice that by definition $\laplaciancoeff^2$ is in general a complex parameter, whose real and imaginary parts are given by
\begin{equation*}
\re\laplaciancoeff^2=\frac{\rpeakphase^2}{6}\frac{\rpeakwidth_r^2-\rpeakwidth_i^2}{\cpeakwidth \combinedpeakparclock^2}\,,\qquad \im\laplaciancoeff^2=\frac{\rpeakphase^2}{3}\frac{\rpeakwidth_r\rpeakwidth_i}{\cpeakwidth \combinedpeakparclock^2}\,.
\end{equation*}
Rewriting explicitly equation \eqref{eqn:redwfunctionevoprio} in terms of these quantities, we thus find
\begin{align}\label{eqn:redwfunctionevo}
0&=\partial^2_0\redwfunction_{\spin}(x,\mommatterfield)-i\firstdercoeff\partial_{0}\redwfunction_{\spin}(x,\mommatterfield)-\realpartnonder_{\spin}^2\redwfunction_{\spin}(x,\mommatterfield)-i\impartnonder_{\spin}^2\redwfunction_{\spin}(x,\mommatterfield)\nonumber\\
&\quad+\re\laplaciancoeff^2\nabla^2\redwfunction_{\spin}(x,\mommatterfield)+i\im\laplaciancoeff^2\nabla^2\redwfunction_{\spin}(x,\mommatterfield)\,,
\end{align}
with
\begin{equation}
\realpartnonder_{\spin}^2\equiv \frac{1}{\cpeakwidth \combinedpeakparclock^2}-\kinratio_{\spin;2}(\mommatterfield)\left(1+3\lambda\re\laplaciancoeff^2\right)\,\qquad \impartnonder_{\spin}^2=3\lambda\im\laplaciancoeff^2 r_{\spin;2}\,.
\end{equation}
This is our fundamental equation determining the form of the reduced condensate wavefunction $\redwfunction$. As in \cite{Oriti:2016qtz,Marchetti:2020umh}, however, it is useful to decompose equation \eqref{eqn:redwfunctionevo} in its real and imaginary parts, by defining $\redwfunction_{\spin}\equiv \rwfunctionmodulus_{\spin}\exp[i\rwfunctionphase_{\spin}]$, so that, using
\begin{align*}
\redwfunction_{\spin}''&=\left[\rwfunctionmodulus_{\spin}''-(\theta'_{\spin})^2\rwfunctionmodulus_{\spin}+i\rwfunctionphase_{\spin}''\rwfunctionmodulus_{\spin}+2i\rwfunctionmodulus_{\spin}'\rwfunctionphase_{\spin}'\right]e^{i\rwfunctionphase_{\spin}}\,,\\
\nabla^2\redwfunction_{\spin}&=\left[\nabla^2\rwfunctionmodulus_{\spin}-(\boldsymbol{\nabla}\rwfunctionphase_{\spin})^2\rwfunctionmodulus_{\spin}+i\nabla^2\rwfunctionphase_{\spin}\rwfunctionmodulus_{\spin}+2i\boldsymbol{\nabla}\rwfunctionmodulus_{\spin}\cdot\boldsymbol{\nabla}\rwfunctionphase_{\spin}\right]e^{i\rwfunctionphase_{\spin}}\,,
\end{align*}
we see that, for the real and imaginary parts we have, respectively,
\begin{subequations}\label{eqn:generalrealimaginary}
\begin{align}
    0&=\rwfunctionmodulus''_{\spin}+\re\laplaciancoeff^2\nabla^2\rwfunctionmodulus_{\spin}-\left[\left(\theta'_{\spin}\right)^2+\realpartnonder_{\spin}^2-\firstdercoeff\rwfunctionphase_{\spin}'-\re\laplaciancoeff^2\left(\boldsymbol{\nabla}\rwfunctionphase_{\spin}\right)^2-\im\laplaciancoeff^2\nabla^2\rwfunctionphase_{\spin}\right]\rwfunctionmodulus_{\spin}\nonumber\\
    &\quad-2\boldsymbol{\nabla}\rwfunctionmodulus_{\spin}\cdot\boldsymbol{\nabla}\rwfunctionphase_{\spin}\,,\\
    0&=\rwfunctionphase_{\spin}''\rwfunctionmodulus_{\spin}+2\rwfunctionphase_{\spin}'\rwfunctionmodulus_{\spin}'-\firstdercoeff\rwfunctionmodulus_{\spin}'+\re\laplaciancoeff^2\left[2\boldsymbol{\nabla}\rwfunctionmodulus_{\spin}\cdot\boldsymbol{\nabla}\rwfunctionphase_{\spin}+\nabla^2\rwfunctionphase_{\spin}\rwfunctionmodulus_{\spin}\right]-\impartnonder_{\spin}^2\rwfunctionmodulus_{\spin}\nonumber\\
&\quad+\im\laplaciancoeff^2\left[\nabla^2\rwfunctionmodulus_{\spin}-\left(\boldsymbol{\nabla}\rwfunctionphase_{\spin}\right)^2\rwfunctionmodulus_{\spin}\right],
\end{align}
\end{subequations}
where we have suppressed the explicit dependence of functions for simplicity. 

At this point, it is important to recall that we are interested in slightly inhomogeneous relational quantities. Therefore, in the next section we will consider a perturbative framework (with respect to spatial gradients) in which we will study the equations above.
\subsection{Background and perturbed equations of motion}\label{subsec:bkgpert}
The perturbative context will be defined by assuming that the functions $\rwfunctionmodulus_{\spin}$ and $\rwfunctionphase_{\spin}$ can be written as
\begin{equation}
    \rwfunctionmodulus_{\spin}=\bkgmodulus_{\spin}+\pertmodulus_{\spin}\,,\qquad\rwfunctionphase_{\spin}\equiv \bkgphase_{\spin}+\delta\rwfunctionphase_{\spin}\,,
\end{equation} 
with $\bkgmodulus=\bkgmodulus(\peakvalue^0,\mommatterfield)$ and $\bkgphase=\bkgphase(\peakvalue^0,\mommatterfield)$ being \virgolette{background} (zeroth-order) quantities and with $\pertmodulus_\spin$ and $\pertphase_\spin$ being small corrections to them. Let us study the zeroth- and the first-order (in $\pertmodulus, \pertphase)$ form of equations \eqref{eqn:generalrealimaginary}.
\paragraph*{Background.}
At the zeroth-order equations \eqref{eqn:generalrealimaginary} become
\begin{align}\label{eqn:backgroundmoduluseq}
\bkgmodulus''_{\spin}(\peakvalue^0,\mommatterfield)-\left[\left(\bkgphase'_{\spin}(\peakvalue^0,\mommatterfield)\right)^2+\realpartnonder_{\spin}^2(\mommatterfield)-\firstdercoeff\bkgphase_{\spin}'(\peakvalue^0,\mommatterfield)\right]\bkgmodulus_{\spin}(\peakvalue^0,\mommatterfield)&=0\,,\\
\bkgphase_{\spin}''(\peakvalue^0,\mommatterfield)\bkgmodulus_{\spin}+2\bkgphase_{\spin}'(\peakvalue^0,\mommatterfield)\bkgmodulus_{\spin}'(\peakvalue^0,\mommatterfield)-\firstdercoeff\bkgmodulus_{\spin}'(\peakvalue^0,\mommatterfield)-\impartnonder_{\spin}^2\bkgmodulus_{\spin}(\peakvalue^0,\mommatterfield)&=0\,,
\end{align}
where we have specified the dependence of the condensate modulus and phase on $\peakvalue^0$ and $\mommatterfield$ explicitly. Let us rewrite the second equation by multiplying by $\bkgmodulus_{\spin}\neq 0$: we obtain
\begin{equation*}
\bkgphase_{\spin}''(\peakvalue^0,\mommatterfield)\bkgmodulus^2(\peakvalue^0,\mommatterfield)_{\spin}+(\bkgphase_{\spin}'(\peakvalue^0,\mommatterfield)-\firstdercoeff/2)(\bkgmodulus_{\spin}^2)'(\peakvalue^0,\mommatterfield)-\impartnonder_{\spin}^2\bkgmodulus^2_{\spin}(\peakvalue^0,\mommatterfield)=0\,, 
\end{equation*}
or, equivalently,
\begin{equation*}
    \bkgphase_{\spin}''(\peakvalue^0,\mommatterfield)+(\bkgphase_{\spin}'(\peakvalue^0,\mommatterfield)-\firstdercoeff/2)\frac{(\bkgmodulus_{\spin}^2)'(\peakvalue^0,\mommatterfield)}{\bkgmodulus_{\spin}^2(\peakvalue^0,\mommatterfield)}-\impartnonder_{\spin}^2=0\,.
\end{equation*}
Now, assume that, in the regime of interest, $\impartnonder_{\spin}^2$ in the above equation is negligible\footnote{\label{footnote:beta}Classically, the volume background dynamics with respect to the scalar field clock is exponential (see Appendix \ref{app:harmonicgauge}). As we will see in Section \ref{sec:evophysicalquantities}, the behavior of the background volume is essentially determined by $\bkgmodulus_\spin^2$. If we make an exponential ansatz (analogous to \eqref{eqn:bkgmodsol}) for $\bkgmodulus_\spin^2$ and we plug it into the above equation for the phase we obtain, for large densities (and thus for large values of the clock, given our exponential ansatz), $\bkgphase'=(\firstdercoeff\massparameter_\spin+\impartnonder_\spin)/(2\massparameter_\spin)$ (assumption \ref{ass:dc1}). By reinserting this into equation \eqref{eqn:backgroundmoduluseq}, we see that the ansatz is not consistent, exactly because of the presence of $\impartnonder_\spin$. This motivates the choice of restricting to small values of $\impartnonder_\spin$. Moreover, as we will see below, since $\impartnonder_{\spin}^2\propto \im \laplaciancoeff$, the regime of small $\impartnonder_\spin$ will be compatible with the decoupling regime for first order perturbations.}. 
The results, in these cases are the same as in \cite{Marchetti:2020umh}, so that the equations for background phase and modulus can equivalently be written in terms of the integration constants $Q_{\spin}$ and $\mathcal{E}_{\spin}$ as 
\begin{subequations}
\begin{align}\label{eqn:bkgphasesol}
\bkgphase'_{\spin}(\peakvalue^0,\mommatterfield)&=\frac{\gamma}{2}+\frac{Q_{\spin}(\mommatterfield)}{\bkgmodulus_{\spin}^2(\peakvalue^0,\mommatterfield)}\mathcomma\\
(\bkgmodulus_{\spin}')^2(\peakvalue^0,\mommatterfield)&=\mathcal{E}_{\spin}(\mommatterfield)-\frac{Q_{\spin}^2(\mommatterfield)}{\bkgmodulus_{\spin}^2(\peakvalue^0,\mommatterfield)}+\massparameter_{\spin}^2(\mommatterfield)\bkgmodulus_{\spin}^2(\peakvalue^0,\mommatterfield)\simeq \massparameter_{\spin}^2(\mommatterfield)\bkgmodulus_{\spin}^2(\peakvalue^0,\mommatterfield)\,,\label{eqn:bkgmodsol}
\end{align}
\end{subequations}
where $\massparameter_{\spin}^2(\mommatterfield)\equiv \realpartnonder_{\spin}^2(\mommatterfield)-\firstdercoeff^2/4$ (we have dropped the superscript $(\lambda)$ for notation simplicity) and with the last approximate equality being valid for large densities $\bkgmodulus_\spin\gg 1$ (assumption \ref{ass:dc1})\label{asspage:dc1a}. 
\paragraph*{First order.}
\label{sec:perturbations}
The first order equations, instead, are
\begin{subequations}\label{eqn:firstorder}
\begin{align}\label{eqn:deltarhofirstorder}
0&=\pertmodulus''_{\spin}(x,\mommatterfield)+\re\laplaciancoeff^2\nabla^2\pertmodulus_{\spin}(x,\mommatterfield)-\realpartnonder_{\spin}^2(\mommatterfield)\pertmodulus_{\spin}(x,\mommatterfield)\nonumber\\
&\quad-\left[\pertphase_{\spin}'(x,\mommatterfield)\left(2\bkgphase'_{\spin}(\peakvalue^0,\mommatterfield)-\firstdercoeff\right)-\im\laplaciancoeff^2\nabla^2\pertphase_{\spin}(x,\mommatterfield)\right]\bkgmodulus_{\spin}(\peakvalue^0,\mommatterfield)\,,\\
0&=\pertphase_{\spin}''(x,\mommatterfield)\bkgmodulus_{\spin}(\peakvalue^0,\mommatterfield)+\bkgphase_{\spin}''(\peakvalue^0,\mommatterfield)\pertmodulus_{\spin}(x,\mommatterfield)+2\pertphase_{\spin}'(x,\mommatterfield)\bkgmodulus_{\spin}'(\peakvalue^0,\mommatterfield)\nonumber\\
&\quad+2\bkgphase_{\spin}'(\peakvalue^0,\mommatterfield)\pertmodulus_{\spin}'(x,\mommatterfield)-\firstdercoeff\pertmodulus_{\spin}'(x,\mommatterfield)+\re\laplaciancoeff^2[\nabla^2\pertphase_{\spin}(x,\mommatterfield)]\bkgmodulus_{\spin}(\peakvalue^0,\mommatterfield)\nonumber\\
&\quad-\impartnonder_{\spin}^2\pertmodulus_{\spin}(x,\mommatterfield)+\im\laplaciancoeff^2\nabla^2\pertmodulus_{\spin}(x,\mommatterfield)\,.\label{eqn:phasefirstorder}
\end{align}
\end{subequations}
The two equations form a complicated set of coupled second order differential equations for the variables $\pertmodulus_{\spin}$ and $\pertphase_{\spin}$. The decoupling regime can be easily identified by first rewriting equation \eqref{eqn:phasefirstorder} as
\begin{align}\label{eqn:phaserewritten}
0&=\bkgmodulus_{\spin}(\peakvalue^0,\mommatterfield)\left[\pertphase_{\spin}''(x,\mommatterfield)+2\pertphase_{\spin}'(x,\mommatterfield)\frac{\bkgmodulus_{\spin}'(\peakvalue^0,\mommatterfield)}{\bkgmodulus_{\spin}(\peakvalue^0,\mommatterfield)}+\re\laplaciancoeff^2\nabla^2\pertphase_{\spin}(x,\mommatterfield)\right]\nonumber\\
&\quad+\pertmodulus_{\spin}(x,\mommatterfield)\left[\bkgphase_{\spin}''(\peakvalue^0,\mommatterfield)+[2\bkgphase_{\spin}'(\peakvalue^0,\mommatterfield)-\firstdercoeff]\frac{\pertmodulus_{\spin}'(x,\mommatterfield)}{\pertmodulus_{\spin}(x,\mommatterfield)}\right]+\im\laplaciancoeff^2\nabla^2\pertmodulus_{\spin}(x,\mommatterfield)\,,
\end{align}
where, similarly to what we did in the background case, and in light of the above discussion, we neglected the term proportional to $\impartnonder_{\spin}^2$. It is then easy to see that the decoupling regime corresponds to the limit in which\footnote{Notice that this condition is consistent with requirement of having negligible $\impartnonder$, see also footnote \ref{footnote:beta}.} (assumption \ref{ass:dc3})\label{asspage:dc3}
\begin{equation}\label{eqn:smallimarginarypartalpha}
\vert \im\laplaciancoeff^2\vert=\frac{2}{3}\frac{\rpeakphase^2\delta_r\vert \delta_i\vert}{\cpeakwidth^2\cpeakphase^2}\ll 1\,,
\end{equation}
and when the background density $\bkgmodulus$ is very large (assumption \ref{ass:dc1})\label{asspage:dc1b}. Indeed, using the background equation \eqref{eqn:bkgphasesol}, equation \eqref{eqn:deltarhofirstorder} can be written as
\begin{equation*}
    L[\pertmodulus_{\spin}]\simeq 2\pertphase_{\spin}'Q_{\spin}/\bkgmodulus_{\spin}\,,
\end{equation*}
with $L$ an appropriate linear differential operator. So, $\pertmodulus\sim \pertphase/\bkgmodulus$, and for large enough $\bar\rho_{\spin}$, the right-hand-side is negligible. Similarly, using that $\bkgphase_{\spin}''=-2Q_{\spin}\bkgmodulus_{\spin}^{-2}(\bkgmodulus_{\spin}'/\bkgmodulus_{\spin})\sim -2Q_{\spin}\massparameter_{\spin}\bkgmodulus_{\spin}^{-2}$, we deduce that the first term at the second line of equation \eqref{eqn:phaserewritten} is of order $\pertmodulus_{\spin}/\bkgmodulus_{\spin}^2$, while the first term at the first line of equation \eqref{eqn:phaserewritten} is of order $\bkgmodulus_{\spin}\pertphase_{\spin}$, so for large enough $\bkgmodulus_{\spin}$, only the latter is important. As a result, equations \eqref{eqn:firstorder} become
\begin{subequations}\label{eqn:decoupledperturbations}
\begin{align}\label{eqn:modulusdecoupled}
    0&\simeq \pertmodulus''_{\spin}(x,\mommatterfield)+\re\laplaciancoeff\nabla^2\pertmodulus_{\spin}(x,\mommatterfield)-\realpartnonder_{\spin}^2(\mommatterfield)\pertmodulus_{\spin}(x,\mommatterfield)\,,\\\label{eqn:thetadecoupled}
    0&\simeq \pertphase_{\spin}''(x,\mommatterfield)+2\pertphase_{\spin}'(x,\mommatterfield)\frac{\bkgmodulus_{\spin}'(\peakvalue^0,\mommatterfield)}{\bkgmodulus_{\spin}(\peakvalue^0,\mommatterfield)}+\re\laplaciancoeff\nabla^2\pertphase_{\spin}(x,\mommatterfield)\,,
\end{align}
\end{subequations}
which are clearly decoupled. An interesting feature of the above equations is that any 
Lorentz property of the second order differential operator in the perturbed equations is in fact only a result of the features of the peaking functions, i.e. of the (approximate) vacuum state we work with, and not of the fundamental symmetries imposed on the GFT action $\gftaction$. Indeed, the parameter $\lambda$, determining whether the matter variables enter the fundamental GFT action in a Lorentz ($\lambda=1$) or Euclidean ($\lambda=-1$) invariant fashion, only enters in $\realpartnonder_{\spin}$, and therefore does not affect at all the differential structure of the equations. Since, as we will see below, the form of the perturbation equations will naturally reflect the structure of equations determining the relational evolution of perturbed physical quantities, this result is particularly intriguing, because it would suggest that only a certain class of states is able to produce relational equations with local Lorentz signature. We will comment further on this in Section \ref{sec:conclusions}.
\section{Effective relational dynamics of physical quantities}\label{sec:evophysicalquantities}
In this section, we will use the evolution equations for the condensate wavefunction in order to obtain relational evolution equations for the expectation values of physical quantities, both at the background, i.e.\ homogeneous, and at the perturbed level, i.e.\ for inhomogeneous cosmological perturbations. In order to keep the notation lighter, for any quantum operator of interest $\op{O}$, we will denote $\bar{O}\equiv \braket{\hat{O}}_{\bar{\wfunction}}$, where the expectation value is computed with respect to the state characterized by the background part of the condensate wavefunction \eqref{eqn:condensatewavefunction}; similarly, we will denote by $\delta O$ the first order term in $\pertmodulus$, $\pertphase$ of the expectation value $\braket{\op{O}}_{\wfunction}$ computed on states characterized by the condensate wavefunction \eqref{eqn:condensatewavefunction}. 

The perturbed relational system includes in general geometric and matter operators. Among the matter operators, those of obvious interest are the $\phi$-scalar field operator and its momentum, written in the $\mommatterfield$ representation (see equations \eqref{eqn:scalarfieldoperator} and \eqref{eqn:momentumoperator}) as
\begin{subequations}\label{eqn:mattervariables}
\begin{align}
\matterfieldop&=\frac{1}{i}\int\diff \gvariables\int\diff^4\framefield\int\diff\mommatterfield\,\gftfieldop^\dagger(\gvariables,\framefield^\mu,\mommatterfield)\partial_{\mommatterfield}\gftfieldop(\gvariables,\framefield^\mu,\mommatterfield)\,,\\
\mattermomop&=\int\diff \gvariables\int\diff^4\framefield\int\diff\mommatterfield\,\mommatterfield\gftfieldop^\dagger(\gvariables,\framefield^\mu,\mommatterfield)\gftfieldop(\gvariables,\framefield^\mu,\mommatterfield)\,.
\end{align}
\end{subequations}
On the geometric side, there are in principle many different operators characterizing the properties of slightly inhomogeneous geometries. Here, we are interested only in scalar perturbations, and in particular only isotropic operators will be considered. Even in this case, however, at the classical level, scalar perturbations are in general captured by several non-trivial functions of the metric components, see e.g.\ equation \eqref{eqn:perturbedlineelement}. Reproducing metric perturbations at the quantum level, however, means determining (i) the structure of microscopic observables and (ii) collective states such that the expectation values of the former on the latter can be associated to emergent metric functions. Most of the work in the literature so far, however, has been devoted to the study of the volume operator  \eqref{eqn:volumeoperator} and to models for which coherent states \eqref{eqn:coherentstates} with wavefunction \eqref{eqn:patchstates} provide an interpretation in terms of metric functions at specific values of the physical frame. 
The definition of more general operators and states is certainly a pressing issue to be tackled in order to define a comprehensive and complete perturbation theory from the GFT framework. However, we will content ourselves with considering the evolution of the universe volume defined (as a quantum operator) in equation \eqref{eqn:volumeoperator}, which is consistent and microscopically well defined, with respect to the states \eqref{eqn:coherentstates} with wavefunction \eqref{eqn:patchstates}.

Moreover, in this section we will consider only the large densities (late times) regime of evolution of the relevant quantities, in which case, as shown in the above section, the equations of motion for $\pertmodulus$ and $\pertphase$ greatly simplify. As explained in Section \ref{sec:condensates}, one would expect this regime (characterized by a very large number of GFT quanta) to be also the classical one (i.e.\ characterized by small quantum fluctuations of macroscopic operators) \cite{Gielen:2019kae,Marchetti:2020qsq}. Therefore, it is of fundamental importance to check whether in this regime the solutions of the equations of motion coming from the quantum theory actually match those of \gr (or possibly of some alternative theory of gravity). This will be the main purpose of the following sections, where geometric (Section \ref{subsec:volumeevolution}) and matter observables\footnote{Here with matter observables we mean the observables associated to the scalar field $\matterfield$, the only relevant contribution to the energy budget of the universe under our assumptions.} (Section \ref{subsec:matterevolution}) will be discussed separately.
More precisely, we will look for a matching with \gr in harmonic gauge (see Appendix \ref{app:harmonicgauge}), which is expected to capture well the physical properties of a relational scalar field frame.

Before going to the actual computations, however, let us mention that the first order perturbed harmonic gauge condition does not produce a complete gauge fixing, meaning that, as we discuss in detail in Appendix \ref{app:harmonicgauge}, there are always small coordinate transformations satisfying \eqref{eqn:residualgaugefreedom} that can be performed while still remaining in the harmonic gauge. An important consequence of this residual gauge freedom is that one can fix the gauge in such a way that volume perturbations are completely absent, by reabsorbing spatial geometric perturbations in the specific coordinate choice. Given the very particular nature of this specific coordinate system, however, it is difficult to understand whether it can actually be realized by a physical reference frame. In particular, since in the quantum theory scalar field operators have matrix elements which are only dependent on \virgolette{pre-matter} variables, one would be led to conclude that using these quantum degrees of freedom as an effective physical reference frame one could not reproduce any such \virgolette{partially geometric} coordinate system. 
Therefore, in the following, we will look for a matching with classical \gr in a completely gauge fixed harmonic gauge, with the understanding that the residual first order gauge freedom has been fixed in such a way that spatial geometric perturbations have not be reabsorbed by the gauge choice.
\subsection{Volume evolution}\label{subsec:volumeevolution}
Let us start from geometric observables, i.e. the evolution of the volume operator. We will first discuss the situation at the homogeneous background level, and then we will move to inhomogeneous perturbations.
\paragraph{Background volume evolution and semi-classical limit.}
The background volume dynamics is given, within our working assumptions and similarly to \cite{Marchetti:2020umh}, by\label{asspage:ks3c}
\begin{equation}\label{eqn:backgrounddynamics}
\left(\frac{\bar{V}'}{\bar{V}}\right)^2\simeq \left(\frac{2\SumInt_{\genlabel}\int\diff\mommatterfield V_{\genlabel}\bkgmodulus_{\genlabel}^2(\peakvalue^0\mommatterfield)\text{sgn}(\bkgmodulus_{\genlabel}'(\peakvalue^0,\mommatterfield))\massparameter_{\genlabel}(\pi_\phi)}{\SumInt_{\genlabel}\int\diff\pi_\phi V_{\genlabel}\bkgmodulus_{\genlabel}^2(\peakvalue^0,\mommatterfield)}\right)^2.
\end{equation}
The classical equations read, instead, in the limit of the field $\phi$ dominating the energy-density budget\footnote{Notice that here we have set the momentum of the clock field to $1$, for simplicity, as it has also been done in Appendix \ref{app:harmonicgauge}. We will discuss below how the results of this and the next section change when the momentum is properly reintroduced.} (see equations \eqref{eqn:friedmanneq} in Appendix \ref{app:harmonicgauge})
\begin{equation}\label{eqn:backgroundclassicaldynamics}
    \left(\frac{\bar{V}'}{\bar{V}}\right)^2=12\pi G(\bar{\pi}^{(c)}_\phi)^2\,,
    \qquad \left[\left(\frac{\bar{V}'}{\bar{V}}\right)^2\right]'=0\,,
\end{equation}
where $\bar{\pi}^{(c)}_{\matterfield}$ is the constant momentum of the scalar fields $\matterfield$, $\bar{\pi}_{\matterfield}^{(c)}=\bar{\matterfield}'$.

To see if equation \eqref{eqn:backgrounddynamics} and its time derivative reduce to equations \eqref{eqn:backgroundclassicaldynamics}, let us consider the case, which has been in fact already shown to have a good classical gravitational interpretation \cite{Marchetti:2020umh,Marchetti:2020qsq, Oriti:2016qtz,Jercher:2021bie}, in which one of the representation labels, say $\sgenlabel$, is dominant (assumption \ref{ass:dc2})\label{asspage:dc2a}. In this case, if $\massparameter_{\sgenlabel}(\mommatterfield)\simeq \gravconst_{\sgenlabel}\mommatterfield$ (assumption \ref{ass:dc4})\label{asspage:dc4}, we have
\begin{equation}\label{eqn:hubbleequation}
\left(\frac{\bar{V}'}{\bar{V}}\right)^2\simeq 4\gravconst_{\sgenlabel}^2\frac{\left[\int\diff\mommatterfield\mommatterfield\bkgmodulus_{\sgenlabel}^2(\peakvalue^0\mommatterfield)\right]^2}{\left[\int\diff\mommatterfield\bkgmodulus_{\sgenlabel}^2(\peakvalue^0,\mommatterfield)\right]^2}=4\gravconst_{\sgenlabel}^2\frac{\bar{\Pi}^2_{\matterfield}}{\bar{N}^2}\,.
\end{equation}
So, when $4\gravconst_{\sgenlabel}^2=12\pi G$, equation \eqref{eqn:backgroundclassicaldynamics} is reproduced by identifying $\bar{\pi}^{(c)}_{\matterfield}\equiv \bar{\Pi}^2_{\matterfield}/\bar{N}^2 $. Notice that for the condition $\massparameter_{\sgenlabel}(\mommatterfield)\simeq \gravconst_{\sgenlabel}\mommatterfield$ to be true, the contribution to $\massparameter_{\sgenlabel}$ due to the geometric coefficients $\kinratio_{\sgenlabel}$ should be dominant, since only they can depend on $\mommatterfield$. In particular, this implies that $\mu_{\sgenlabel}\simeq \realpartnonder_{\sgenlabel}$, since they only differ by a $\mommatterfield$-independent coefficient.

However, while the above conditions are clearly sufficient to reproduce the first Friedmann equation, they are not in general enough to guarantee the validity of the second Friedmann equation, stating that $(\bar{V}'/\bar{V})'=0$. The reason for this is that the ratio $\bar{\Pi}_{\matterfield}/\bar{N}$ is in general \emph{not constant}:
\begin{equation*}
\left[\frac{\bar{\Pi}_{\matterfield}}{\bar{N}}\right]'=2\left[\frac{\int\diff\mommatterfield\mommatterfield\massparameter_{\sgenlabel}(\mommatterfield)\bkgmodulus_{\sgenlabel}^2(\peakvalue^0,\mommatterfield)}{\int\diff\mommatterfield\bkgmodulus_{\sgenlabel}^2(\peakvalue^0,\mommatterfield)}-\frac{\left[\int\diff\mommatterfield\mommatterfield\bkgmodulus_{\sgenlabel}^2(\peakvalue^0,\mommatterfield)\right]\left[\int\diff\mommatterfield\massparameter_{\sgenlabel}(\mommatterfield)\bkgmodulus_{\sgenlabel}^2(\peakvalue^0,\mommatterfield)\right]}{\left[\int\diff\mommatterfield\bkgmodulus_{\sgenlabel}^2(\peakvalue^0,\mommatterfield)\right]^2}\right].
\end{equation*}
If we assume, as before, that $\massparameter_{\genlabel}\simeq \gravconst_{\genlabel}\mommatterfield$, we see that the right-hand-side of the above equation has the form $\bar{\Pi}_{\phi,2}/\bar{N}-\bar{\Pi}_{\phi}^2/\bar{N}^2$, where $\bar{\Pi}_{\phi,2}$ is the background expectation value of the second quantized operator $\framemomop_{\phi,2}$ whose matrix elements in momentum space are given by $\mommatterfield^2$. In general, this quantity is not zero. However, if we further assume, as done in \cite{Gielen2020} that the condensate wavefunction has a peaking part peaked on one value of the momentum, say $\peakvaluemom$ of $\phi$, so that the condensate wavefunction can be written as\footnote{Notice that changing the form of the condensate wavefunction from equation \eqref{eqn:condensatewavefunction} to \eqref{eqn:finalcondensatewavefunction} and by assuming that $\mompeakingfunc$ is independent on the clock variables (as we are doing here) does not affect the equations of motion of $\rwfunctionmodulus_{\mompeakingwidth}$ and $\rwfunctionphase$ at all because of their linearity.} (assumption \ref{ass:kc3})\label{asspage:kc3a}
\begin{equation}\label{eqn:finalcondensatewavefunction}
\wfunction_{\cpeakwidth,\delta,\cpeakphase,\rpeakphase;\peakvalue^\mu;\peakvaluemom}=\peakfunc_\cpeakwidth(\framefield^0-\peakvalue^0;\cpeakphase)\peakfunc_{\delta}(\vert\framefieldbf-\mathbf{x}\vert;\rpeakphase)\mompeakingfunc_{\mompeakingwidth}(\mommatterfield-\peakvaluemom)\redwfunction(\gvariables,\framefield^0,\framefieldbf,\mommatterfield)\,,
\end{equation}
we find that $\bar{\Pi}_{\phi,2}/\bar{N}-\bar{\Pi}_{\phi}^2/\bar{N}^2\simeq \peakvaluemom^2-\peakvaluemom^2=0$, and both Friedmann equations are thus satisfied, giving
\begin{equation}
    \mathcal{H}^2\equiv \left(\frac{\bar{V}'}{3\bar{V}}\right)^2=\frac{4}{9}\massparameter^2_{\sgenlabel}(\peakvaluemom)=\frac{4\pi G}{3} \peakvaluemom^2\,,\qquad \mathcal{H}'=0\,.
\end{equation}
This also leads to the interpretation of $\peakvaluemom$ as the background classical momentum of the scalar field $\phi$, $\bar{\pi}^{(c)}_\phi$. We will discuss this point in more detail in the next section. 

Finally, let us emphasize that the above peaking condition on $\peakvaluemom$ is not related to an implementation of effective relational localization. However, it does localize the wavefunction in \virgolette{momentum space}, since, as we have mentioned, $\peakvaluemom$ can be identified with $\bar{\pi}^{(c)}_\phi$. In turn, since $\bar{\pi}^{(c)}_\phi$  is the quantity appearing at the right-hand-side of the classical Friedmann equations, it may well be that this localization property is associated to some form of semi-classicality. We leave further investigations of the physical interpretation of this peaking property to future works.
\paragraph{Perturbed volume evolution.}
As before, let us assume that we are in the case of single representation label dominance (assumption \ref{ass:dc2}). Then, the average perturbed volume reads
\begin{equation}
    \delta V(\peakvalue,\peakvaluemom)
    \simeq 2V_{\genlabel_o}\bkgmodulus_{\genlabel_o}(\peakvalue^0,\peakvaluemom)\delta\bkgmodulus_{\sgenlabel}(x,\peakvaluemom)\,,
\end{equation}
where we have used the peaking properties in $\mommatterfield$ of the condensate wavefunction \eqref{eqn:finalcondensatewavefunction}. Now, let us take a time derivative of the above quantity. We have
\begin{align*}
    \delta V'(\peakvalue,\peakvaluemom)&=2V_{\sgenlabel}\bkgmodulus_{\sgenlabel}'(\peakvalue^0,\peakvaluemom)\pertmodulus_{\sgenlabel}(x,\peakvaluemom)+2V_{\sgenlabel}\bkgmodulus_{\sgenlabel}(\peakvalue^0,\peakvaluemom)\pertmodulus'_{\sgenlabel}(x,\peakvaluemom)\\
    &\simeq \massparameter_{\sgenlabel}(\peakvaluemom)\delta V(\peakvalue,\peakvaluemom)+2V_{\sgenlabel}\bkgmodulus_{\sgenlabel}(\peakvalue^0,\peakvaluemom)\pertmodulus'_{\sgenlabel}(x,\peakvaluemom)\,,
\end{align*}
where in the second line we have used the large $\bkgmodulus_{\sgenlabel}$ behavior\footnote{\label{footnote:signmu}For concreteness, we are considering large positive times $\peakvalue^0$, so that only the positive root of equation \eqref{eqn:bkgmodsol} is important.} $\bkgmodulus_{\sgenlabel}'\simeq \massparameter_{\sgenlabel}\bkgmodulus_{\sgenlabel}$. Taking one further derivative and using the above equation together with \eqref{eqn:modulusdecoupled}, we find
\begin{equation}
    \delta V''-2\massparameter_{\sgenlabel}\delta V'+\re\laplaciancoeff\nabla^2\delta V+\delta V(\realpartnonder_{j}^2-\massparameter_j^2)=0\,.
\end{equation}
Recall also that, by consistency with the background equations, we have that $\massparameter_j^2\simeq \realpartnonder_{j}^2$, thus leading to the simplified form 
\begin{equation}\label{eqn:pertvolumegeneralform}
    \delta V''-2\massparameter_{\sgenlabel}\delta V'+\re\laplaciancoeff\nabla^2\delta V=\delta V''-3\mathcal{H}\delta V'+\re\laplaciancoeff\nabla^2\delta V=0\,.
\end{equation}
From the above equation we notice, in particular, that in order to have a Lorentz signature for the equation of physical perturbations we need to require $\re\laplaciancoeff<0$, which in turn implies $\delta_i^2>\delta_r^2$ (see assumption \ref{ass:kc1}\label{asspage:kc1b}). Moreover, in the extreme case in which $\delta_i^2\gg \delta_r^2$, and when $\rpeakphase^2\delta_i^2=3\cpeakwidth^2\cpeakphase^2$ (in which case the parameters of the peaking functions are chosen of the same magnitude), we have 
\begin{equation}\label{eqn:lorentz}
\re\laplaciancoeff^2=\frac{\rpeakphase^2}{6\cpeakwidth \combinedpeakparclock^2}\left(\delta_r^2-\delta_i^2\right)\simeq -\frac{\rpeakphase^2\delta_i^2}{3\cpeakwidth^2\cpeakphase^2}=-1\,,
\end{equation}
which in turn implies that the second order differential operator appearing in \eqref{eqn:pertvolumegeneralform} can be recast in terms of a $\Box$ operator.

By comparing equations \eqref{eqn:pertvolumegeneralform} and \eqref{eqn:classicalvolumeequation}, however, we conclude that the effective evolution of the perturbed volume obtained from our quantum gravity model \emph{does not match} the classical \gr one, in general. An important difference lies in the pre-factor of the Laplacian term of the equation\footnote{Notice, however, that the general spatial differential structure of the equations is the same, thus implying that in the limit of $k\to\infty$ (with all the remaining quantities kept constant), the two equations are equivalent.}, being (in general) respectively $\re\laplaciancoeff$ and $\propto \bar{V}^{4/3}$ in equations \eqref{eqn:pertvolumegeneralform} and \eqref{eqn:classicalvolumeequation}.  We will comment on the possible implications of this mismatch in Section \ref{sec:conclusions}.

In the super-horizon limit $k\to 0$ (where $k$ represents the modulus of the modes associated to a spatial Fourier transform of the perturbed volume), thus for long-wavelength perturbations, equation \eqref{eqn:pertvolumegeneralform} admits two solutions: a constant one, and one of the form $\delta V\propto \bar{V}$. The latter becomes dominant as the universe expands, i.e. at large universe volumes. From the results in Appendix \ref{app:harmonicgauge} (see equation \eqref{eqn:limitspsi} and the discussion below equation \eqref{eqn:classicalvolumeequation}), we see that this dominant solution actually coincide with the \gr one in the limit $k\to 0$. Thus, we conclude that the theory matches the predicted dynamics of \gr in the super-horizon regime, at late cosmological times and large universe volume (which is also when the background dynamics reproduces the Friedmann one). 

\subsection{Matter evolution}\label{subsec:matterevolution}\label{asspage:ks3d}\label{asspage:kc3b}\label{asspage:dc2b}
Let us now move to matter variables, i.e. to the background and perturbed expectation values of the operators $\matterfieldop$ and $\op{\Pi}_\phi$ defined in \eqref{eqn:mattervariables}. Their expectation values read, respectively\footnote{Here, for notational simplicity, we have reabsorbed any phase of the peaking function $\mompeakingfunc_{\mompeakingwidth}\equiv \vert \mompeakingfunc_{\mompeakingwidth}\vert e^{i\theta_f}$ into the phase of the reduced condensate wavefunction, redefining the global phase factor $\theta_{\sgenlabel}$.}
\begin{subequations}\label{eqn:generalmatter}
\begin{align}
\braket{\matterfieldop}_{\wfunction}&\simeq \rho^2_{\sgenlabel}(\peakvalue,\peakvaluemom)[\partial_{\mommatterfield}\theta_{\sgenlabel}](\peakvalue,\peakvaluemom)=[\partial_{\mommatterfield}\theta_{\sgenlabel}](\peakvalue,\peakvaluemom) N(\peakvalue,\peakvaluemom)\,,\\
\braket{\mattermomop}_{\wfunction}&\simeq \peakvaluemom\rho^2_{\sgenlabel}(\peakvalue,\peakvaluemom)= \peakvaluemom N(\peakvalue,\peakvaluemom)\,.
\end{align}
\end{subequations}
As we did for the volume operator, let us write explicitly the contributions to these quantities at the background and perturbed level.
\paragraph*{Background.}
At the background level from equations \eqref{eqn:generalmatter}, we have
\begin{equation*}
    \bar{\Pi}_\phi\simeq \peakvaluemom\bar{N}(\peakvalue^0,\peakvaluemom)\,,\qquad  \bar{\Phi}\simeq \bar{N}(\peakvalue^0,\peakvaluemom)[\partial_{\mommatterfield}\bkgphase_{\sgenlabel}](\peakvalue^0,\peakvaluemom)\,.
\end{equation*}
The dynamics of the background phase $\bkgphase_{\sgenlabel}$ is determined by equation \eqref{eqn:bkgmodsol}:
\begin{equation}\label{eqn:backgroundphase}
\bkgphase_{\sgenlabel}'=\frac{\firstdercoeff}{2}+\frac{Q_{\sgenlabel}}{\bar\rho_{\sgenlabel}^2}\,,\qquad\text{where}\qquad (\bkgmodulus_{\sgenlabel}^2)'\simeq \massparameter_{\sgenlabel}^2\bkgmodulus^2_{\sgenlabel}\,,
\end{equation}
with a prime denoting as usual a derivative with respect to the scalar time\footnote{In the equation for $\bkgmodulus_{\sgenlabel}^2$ we have neglected lower order terms in powers of $\bar{\bkgmodulus}_{\sgenlabel}^2$, since in the above equation for $\bkgphase_{\sgenlabel}$ we are already considering contributions suppressed as $\bkgmodulus^{-2}_{\sgenlabel}$. Any correction to the second equation in \eqref{eqn:backgroundphase} would thus result in even more negligible contributions to the first equation of \eqref{eqn:backgroundphase}.}. Integrating the equation on the left using the equation on the right we obtain
\begin{equation}
\bkgphase_{\sgenlabel}=\frac{\firstdercoeff}{2}\peakvalue^0-\frac{Q_{\sgenlabel}}{\massparameter_{\sgenlabel}\bar\rho_{\sgenlabel}^2}+C_{\sgenlabel}\,,
\end{equation}
where $C_{\sgenlabel}$ is an integration constant and where we have chosen a specific root for the second equation in \eqref{eqn:backgroundphase} (see footnote \ref{footnote:signmu}). Now, it is important to notice that $\firstdercoeff$ does not depend on $\mommatterfield$, while $\mu$ and $Q$ (and $C$) in principle do, even if they do not depend on time. As a result, we have, 
\begin{equation}
\bar{\Phi}\simeq \left[-\partial_{\mommatterfield}\left[\frac{Q_{\sgenlabel}}{\massparameter_{\sgenlabel}}\right]+Q_{\sgenlabel}\frac{\partial_{\mommatterfield}\massparameter_{\sgenlabel}}{\massparameter_{\sgenlabel}}\peakvalue^0+\bar{N}\partial_{\mommatterfield}C_{\sgenlabel}\right]_{\mommatterfield=\peakvaluemom}\,.
\end{equation}
These results should be compared with the classical dynamics, given by $\bar{\phi}''=0$, i.e., $\bar{\phi}=c_1+c_2\peakvalue^0$, and of course also implying that the momentum of $\bar{\phi}$, $\bar{\pi}^{(c)}_\phi$, is a constant. Given the presence of $\bar{N}$ in the above expectation values for $\framemomop$ and $\matterfieldop$ (which grows exponentially in relational time), it is clear that we can only define 
\begin{equation}\label{eqn:scalarfieldmomentumbackground}
    \bar{\pi}^{(c)}_\phi\equiv \bar{\Pi}/\bar{N}=\peakvaluemom\,,
\end{equation}
with $\peakvaluemom$ that would be then associated to the classical momentum of the scalar field, $\peakvaluemom=\bar{\pi}^{(c)}_\phi$, which is the same identification we have found in the previous section by comparing the quantum volume evolution equations with the classical one. Notice that as a consequence of equation \eqref{eqn:scalarfieldbackground} we would also expect $\bar{\phi}'\equiv \bar{\pi}^{(c)}_\phi=\peakvaluemom$. 

The same reasoning is not adequate, instead, for the massless scalar field operator. Indeed, for large $\bar{N}$, a constant term (independent on the scalar field clock) becomes dominant, meaning that by defining $\phi=\braket{\matterfieldop}_{\wfunction}/\bar{N}$, we cannot match the classical result. On the other hand, if we take $C_{\sgenlabel}$ to be independent on $\mommatterfield$, $\braket{\matterfieldop}_{\wfunction}$ becomes an intensive quantity that can be readily compared to $\bar{\phi}$. In this case, consistency with the momentum correspondence requires that $[Q_{\sgenlabel}\partial_{\mommatterfield}(\log \massparameter_{\sgenlabel})]_{\mommatterfield=\peakvaluemom}=\peakvaluemom$. By assuming, as we did to arrive to equation \eqref{eqn:hubbleequation}, that $\massparameter_{\sgenlabel}\simeq \gravconst_{\sgenlabel}\mommatterfield$, the above condition fixes $Q_{\sgenlabel}\simeq \mommatterfield^2$, and, as a result,
\begin{equation}\label{eqn:scalarfieldbackground}
\bar{\phi}\equiv \braket{\matterfieldop}_{\wfunction}\simeq -\gravconst_{\sgenlabel}^{-1}+\peakvaluemom\peakvalue^0\,.
\end{equation}
Notice that, in \eqref{eqn:hubbleequation}, the quantity $\gravconst_{\sgenlabel}$ fixes the gravitational constant. This means that, for a finite value of the gravitational constant, one can never have $\gravconst_{\sgenlabel}^{-1}=0$. This implies that the background matter field \emph{can never be identified with the clock field}. One could argue that this condition is in fact necessary in order to trust the effective relational framework we have defined. It is intriguing, however, that this fact is related to the the gravitational constant attaining a finite value. Seen the other way around, it is interesting that the gravitational constant is in fact determined by the specific matter content that is present in the underlying fundamental theory. Both these points certainly deserve further scrutiny.

Let us conclude this discussion with two comments.
\begin{itemize}
    \item First, we emphasize that the matching with the classical equations has been performed by choosing the specific classical gauge fixing defined by the lapse function $N=a^3$, which naturally fixes the classical momentum of the scalar field as $\clockmomclassic=1$. Of course, one could have fixed $\clockmomclassic$ to any other arbitrary constant, in which case, at the classical level, one would have $\bkgmatter'=\bkgmattermomclassic/\clockmomclassic$. As a consequence, therefore, the matching would have required, e.g., $Q_{\sgenlabel}=\mommatterfield^2/\clockmomclassic$, and $c_j^2=3\pi G/\clockmomclassic$. However, requiring the quantity $\clockmomclassic$ to be in direct correspondence with the expectation value of the clock field momentum would be difficult to justify from a physical perspective. Indeed, the classical momentum is always defined as a result of an (arbitrary) choice of time coordinate, which in the fundamental quantum gravity theory is simply absent. We therefore refrain from requiring any such correspondence, and we consider, from now on, the case $\clockmomclassic=1$ in order make the matching with the classical theory more straightforward.
    \item It is also important to emphasize that the definition of $\bar{\Phi}$ and $\bar{\Pi}_\phi$ in terms of the fundamental operators is different from the one that is used for the clock variables in \cite{Marchetti:2020umh}. Indeed, in \cite{Marchetti:2020umh}, the value of the scalar field clock was associated to the scalar clock field operator divided by the particle number, while the momentum was just associated to the clock momentum operator. The tension between the definitions provided in \eqref{eqn:scalarfieldbackground} and \eqref{eqn:scalarfieldmomentumbackground} may be solved by noticing that the definitions are in fact consistent when one considers how the associated variables are represented. Indeed, for the clock scalar field, we have chosen the coordinate representation, and thus divided by $\bar{N}$ the operator which is multiplicative in this representation. The same has been done for the source scalar field, but since in this case the representation used is the momentum one, it has been the momentum operator (diagonal in this representation) that has been divided by $\bar{N}$. On the other hand, operators with derivative kernels in the two representations needed not to be divided by $\bar{N}$. While in principle the two representations are absolutely equivalent, the kind of states we have chosen, and according to which the definitions of $\bar{\phi}$ and $\bar{\pi}_\phi$ are provided, clearly distinguishes between the different sets of variables.
\end{itemize}
\paragraph*{Perturbed scalar field evolution.}
Similarly to what we did for the volume operator, we can study perturbations to the scalar field quantities. Notice, however, that results about perturbations on the matter sector depend on how extensive variables are matched with classical ones. For instance, for the second quantized field operator, we have seen that $\phi=\braket{\matterfieldop}_{\wfunction}$, so
\begin{equation}
    \pertmatter=\delta\braket{\matterfieldop}_{\wfunction}=\left[\frac{\delta N}{\bar{N}}\bkgmatter+\bar{N}\partial_{\mommatterfield}\pertphase_{\sgenlabel}\right]_{\mommatterfield=\peakvaluemom}\,.
\end{equation}
The dynamical equation satisfied by $\pertmatter$ can be easily determined by noticing that $\delta N/\bar{N}=2\pertmodulus_{\sgenlabel}/\bkgmodulus_{\sgenlabel}\equiv 2\delta_{\rwfunctionmodulus_{\sgenlabel}}$, and that $\delta_{\rwfunctionmodulus_{\sgenlabel}}$ and $\pertphase_{\sgenlabel}$ satisfy the same differential equation:
\begin{equation}
    \delta_{\rwfunctionmodulus_{\sgenlabel}}''+2\mu_{\sgenlabel}\delta_{\rwfunctionmodulus_{\sgenlabel}}'+\re\laplaciancoeff\nabla^2\delta_{\rwfunctionmodulus_{\sgenlabel}}=0=\pertphase_{\sgenlabel}''+2\mu_{\sgenlabel}\pertphase_{\sgenlabel}'+\re\laplaciancoeff\nabla^2\pertphase_{\sgenlabel}\,.
\end{equation}
Here, consistently to what was done in the previous sections, we have used $\bkgmodulus'_{\sgenlabel}\simeq \mu_{\sgenlabel}\bkgmodulus$ and the fact that $\mu_{\sgenlabel}\simeq \realpartnonder_{\sgenlabel}$. 

Already from this equation, in particular from the behavior of the spatial derivative term (scaling as $\bar{V}^{4/3}$, see equation \eqref{eqn:deltaphiharmonicgauge}) we can conclude that the evolution equation for the scalar field perturbations does not match, in general, with the \gr one. 

Still, similarly to what happens for the volume perturbations, we can verify that solutions to our equations and to the \gr ones do match in the super-horizon regime (long wavelength limit). To see this explicitly, notice that in this case the equation satisfied by $\pertphase_{\sgenlabel}$ becomes
\begin{equation}\label{eqn:deltathetalargewave}
0=\pertphase_{\sgenlabel}''+\frac{(\bkgmodulus_{\sgenlabel}^2)'}{\bkgmodulus_{\sgenlabel}^2}\pertphase'_{\sgenlabel}=\pertphase_{\sgenlabel}''+\massparameter_{\sgenlabel}\pertphase'_{\sgenlabel}\,,\qquad \superhor\,,
\end{equation}
whose general solution is
\begin{equation}
\pertphase_{\sgenlabel}=c_{1,\sgenlabel}(\mommatterfield)+c_{2,\sgenlabel}(\mommatterfield)\bar{N}^{-1}\,,\qquad \superhor\,,
\end{equation}
with an appropriate redefinition of constants. Thus, in the large $\bar{N}$ limit, we can write $\pertphase_{\sgenlabel}\simeq c_{1,\sgenlabel}(\mommatterfield)$, and since $\delta N/\bar{N}$ is constant, $\delta\phi\simeq \bar{N}c_{1,\sgenlabel}(\mommatterfield)$, which forces us to consider $c_{1,\sgenlabel}$ to be independent on $\mommatterfield$ in order to match with \gr. Indeed, in this case, we have
\begin{equation}
\delta \phi=\delta\braket{\matterfieldop}=\left[\frac{\delta V}{\bar{V}}\bar{\phi}+\partial_{\mommatterfield}c_{2,\sgenlabel}-c_{2,\sgenlabel}\partial_{\mommatterfield}\massparameter_{\sgenlabel}\peakvalue^0\right]_{\mommatterfield=\peakvaluemom} \,,\qquad \superhor\,,
\end{equation}
which is compatible with the classical solution, since in virtue of $\delta V/\bar{V}$ being constant\footnote{Recall that the dominant solution of equation \eqref{eqn:pertvolumegeneralform} in the $k\to 0$ limit is $\delta V\propto \bar{V}$.}, it satisfies $\delta\phi''=0$. 

Let us now consider perturbations in the scalar field momentum. If the classical momentum $\pi_{\phi}^{(c)}$ is identified with $\braket{\framemomop_{\phi}}_{\wfunction}/N\simeq \peakvaluemom$, we have that $\delta \pi_\phi=0$. On the other hand, if we maintain the correspondence suggested above, i.e.\ $\pi_\phi=\braket{\framemomop_{\phi}}_{\wfunction}/\bar{N}$, then  
\begin{equation}
    \delta \pi_\phi=\peakvaluemom\frac{\delta N}{\bar{N}}=\peakvaluemom\frac{\delta V}{\bar{V}}\,,
\end{equation}
with the equations for $\delta V$ being already described in the previous subsection. In order to have a consistent definition of the momentum, however, we should require $\delta\matterfield'=\delta\mommatterfield$, which, in the long wavelength limit forces us to impose $c_{2,\sgenlabel}=0$, so that  we find
\begin{equation}
    \pertmatter=(\delta V/\bar{V})\bkgmatter\,\,,\qquad \superhor\,.
\end{equation}
Therefore we see that in the super-horizon limit perturbations are only present in the modulus of the condensate wavefunction. 

In conclusion, there is no matching with the classical theory for arbitrary wavelengths, and we obtain instead a modified dynamics for cosmological perturbations, from our quantum gravity model. On the other hand, we see that the same assumptions needed for the background solutions to match \gr allow also for a classical matching of perturbed quantities in the super-horizon limit (which is expected already from previous work on this issue in the separate universe framework). This is a good consistency check of our formalism and procedure. Still, we have seen that the discussion of inhomogeneous perturbations is complicated by the fact that one needs to identify also the right way to turn extensive quantities into intensive ones (i.e., $\bar{N}$ or $N$, as we have seen in the case of the perturbed momentum). Consistent and rather compelling choices can be identified, though. We remark that this additional difficulty is due to the fact that our fundamental degrees of freedom are not quantized fields and that spatiotemporal observables emerge, in this \qg formalism, only as collective, averaged quantities. This is reflected in the presence of an additional observable, with no classical counterpart, given by the number operator. It is the correct way of using this additional, purely quantum gravity observable, that needs to be determined, in order to match continuum gravitational physics.

\section{Summary and discussion}\label{sec:conclusions}
In this section we provide a summary and a discussion of the main results presented in the paper (Section \ref{sec:summary}). Moreover, we review the approximations and assumptions (and the arguments motivating them) made in order to obtain such results (Section \ref{sec:approx}), in particular discussing how relaxing some of them may impact the final results and help recover the appropriate \gr limit for arbitrary wavelengths.
\subsection{Summary and outlook}\label{sec:summary}
In this paper we have studied the extraction of scalar (and isotropic) cosmological perturbations from full \qg, within the \GFT formalism. The classical counterpart of the \GFT system we have considered consists of five massless and free scalar fields minimally coupled to geometry, four of which constitute the material reference system and which by assumption provide a negligible contribution to the total energy-momentum budget of the \universe. The remaining field (called $\matterfield$) is assumed to be slightly inhomogeneous with respect to the matter reference frame.

In contrast to past works on the subject \cite{Gielen:2017eco,Gielen:2018xph,Gerhardt:2018byq}, here we have for the first time defined perturbations (of matter and geometric quantities) in an effective relational sense, by means of  coherent states sharply peaked on values associated to the four massless scalar fields (whose interpretation stems from the role their corresponding discrete degrees of freedom play in the fundamental perturbative quantum dynamics of the \GFT models) we used as a matter frame. These peaking values are then the quantities with respect to which other physical observables can be relationally localized. Being this relational localization by construction only effective, controlled both by the above peaking properties of the condensate wavefunction and by the averaged total number of microscopic \GFT quanta (controlling also the magnitude of quantum fluctuations), it bypasses most of the technical difficulties associated to a material frame composed by four minimally coupled scalar fields highlighted e.g.\ in \cite{Giesel:2016gxq}.

By imposing an averaged form of the quantum many-body microscopic \GFT dynamics, we have obtained dynamical equations determining the phase and the modulus of the (reduced) condensate wavefunction. These phase and modulus, representing the basic hydrodynamic variables of our \virgolette{quantum spacetime fluid}, are then assumed to split into background quantities taken to be homogeneous (i.e.\ independent on the relational rods) and small inhomogeneous perturbations, which allowed in turn for a perturbative analysis of the condensate equations. The resulting first order equations for phase and modulus are in general coupled, but one can show that they actually decouple in the limit of large average number of \GFT quanta, a limit which has been associated to the emergent classical behavior of the macroscopic spacetime quantities \cite{Marchetti:2020qsq, Gielen:2019kae}. 

We have also seen that the equation for the perturbed modulus, eventually determining the behavior of crucial quantities, like e.g.\ the perturbed volume operator, shows an effective Lorentz signature of the derivative kernel only if one assumes that the width of the peaking condensate function (assumed to be isotropic for simplicity) on the scalar field rods is in general complex, with a large imaginary part (and a positive real one, guaranteeing in fact the aforementioned peaking properties). Interestingly enough, this feature seems to be independent on the symmetry properties of the classical action of the scalar fields, in turn assumed to be respected by the \GFT action $\gftaction$, but to really only depend on the dispersion around the peaking value of the peaking states. How this emergent Lorentz signature is related to the local Lorentz structure encoded in the group data associated to the \GFT field is still an open and fundamental question.

The decoupled dynamics (at large universe volumes) of perturbed phase and modulus of the condensate wavefunction has then been used in order to study the dynamics of macroscopic observables associated to geometry (i.e.\ the volume operator) and to the non-reference matter scalar field $\matterfield$, in order to match it with \gr (with a first order harmonic gauge fixing). The background, unperturbed dynamics turns out to be consistent with \gr in the limit of one dominating representation label and with the additional assumption that the condensate wavefunction also peaks on an arbitrary value $\peakvaluemom$, of the variable conjugate to the field $\matterfield$.

The background analysis of the equations shows two more interesting properties. 

First, one is naturally led to associate the classical scalar field $\matterfield$ with the \emph{extensive} (second quantized) scalar field operator associated to $\matterfield$, contrarily to what is actually done for the reference fields. This would suggest that the determination of classical quantities in terms of intrinsic and extensive expectation values of second quantized ones, actually depends on the physical role the quantities themselves play. 

Second, the shift between the clock scalar field and the background part of $\phi$ enters directly in the emergent definition of the gravitational constant. This is intriguing not only because it implies that the matter content of the \universe actually impacts the very definition of the emergent gravitational constant, but it also highlights a possible interesting connection with the very notion of relational evolution. Indeed, a non-zero shift, implying that one can never identify the clock field with the background matter field, in turn means that the gravitational constant cannot vanish. It will be interesting to investigate further the interplay between relationality and emergent constant of nature by considering how to switch between two similar clocks, an issue which, by itself, is likely to play a crucial role in answering open questions about the emergence of relational dynamics from \qg (including, e.g.\ the role that unitarity plays in the fundamental theory). 

Finally, we have studied the dynamics of linear perturbations in volume and matter variables. Obtaining such explicit dynamical equations directly from the full quantum gravity theory, within several approximations, which are however under control (at least in principle) within the fundamental formalism, is our main result. 

For both types of quantities, the solutions to the equations of motion actually match \gr in the super-horizon limit of very large wavelengths. However, in other regimes, starting already at intermediate wavelengths, before entering the sub-horizon regime, \emph{no such match} was found (though the classical and quantum equations do share the same spatial differential structure, characterized by a Laplace operator on the rod fields). 

There are three ways in which we could interpret this result. 

First, we could insist that in this regime the dynamics of cosmological perturbations, as derived from quantum gravity, should in fact match the \gr one. This implies that some assumptions that were taken in our derivation were not justified, and need to be improved (see Section \ref{sec:approx}). Of course, we know well that such improvements are needed independently of this issue, but it is worth considering which ones could be responsible for this specific mismatch. These could involve, for instance, the peaking parameters which were taken to be actually independent on geometric data (assumption \ref{ass:ks3}), or the \GFT interactions being entirely neglected in deriving the equation for cosmological perturbations\footnote{It is worth noticing however that it is the mesoscopic regime of free \GFT dynamics at large volumes, before interactions become relevant, that gives an effective Friedmann dynamics for the background} (assumption \ref{ass:ds3}). 

Second, the identification of perturbations we used may be incorrect or at least insufficient to reproduce their correct dynamics. We have indeed assumed \emph{condensate perturbations}, but it is easy to envisage situations (which are actually of great physical interest \cite{pitaevskii2016bose} in condensed matter physics) where fluid perturbations are determined also by small components of the fluid out of the condensate, and that of course a better approximation of the true condensate (vacuum) state, beyond the mean-field or the perfect condensate (or coherent) state should be considered (see assumption \ref{ass:ds2}). 

Finally, it is possible that none of the two interpretations above is correct, meaning that the procedure we employed is sufficient to extract the relevant effective continuum dynamics of cosmological perturbations from \qg. In other words, our \qg theory, that is our chosen class of \GFT models, simply predicts the above mismatch with \gr. Of course, this would be the most intriguing possibility.  
Taking this possibility seriously, we have again two possible implications. First, the underlying \GFT (and spin foam) models are \virgolette{lacking} and we managed to identify a problematic aspect of them in terms of their physical predictions (e.g.\ within assumption \ref{ass:ds1}). This in itself would be quite an achievement, in our opinion, given how difficult it is to constrain quantum gravity models in general. Second, the modified dynamics of cosmological perturbations we have obtained should be taken seriously as a prediction and it may turn out to be in fact a viable one. For example, it could correspond to the predicted dynamics of some modified gravity theory, which would then be the effective continuum and classical description of our fundamental quantum gravity dynamics\footnote{Let us point out that the effective cosmological dynamics for the background, obtained from these \GFT models, has been already matched with (limiting curvature) mimetic gravity \cite{deCesare:2018cts}.}. In order to explore this possibility, we need to invest serious work on the comparison of these predictions with observational cosmological data, and it would be exciting work.

In conclusion, in this work we have taken one important step towards the definition of a complete framework for the extraction of an effective relational cosmological perturbation theory from full \qg. The results of the analysis showed perturbative consistency with \gr (in first order harmonic gauge) in the super-horizon limit, but a modified dynamics of cosmological perturbations otherwise. Our analysis has allowed us to gain some important new insights on the emergence of a relational Lorentz-like structure in the macroscopic dynamics, and to further characterize the emergent physics of \GFT models including not only (four) reference fields, but also additional, non-frame matter. Moreover, this work has highlighted what steps are likely to be crucial for the ambitious goal of a complete extraction of cosmological perturbation theory from \qg. Among them, we emphasize (i) the need for the construction of microscopic observables whose expectation value on appropriate states can be associated to macroscopic geometric quantities (ideally one would aim for the construction of an emergent metric); (ii) the need for an in-depth investigation on the relation between the Lorentz-like properties of the emergent dynamics and those encoded in the group theoretic data assigned to the microscopic \GFT field; and (iii) the study of the impact of \virgolette{out-of-condensate} perturbations on the macroscopic emergent dynamics and, more generally, improvements of the various approximations that were needed to get to the results obtained in this work.

\subsection{Approximations and assumptions}\label{sec:approx}

The assumptions made in this paper can be naturally split in kinematic and dynamic ones; moreover, we will also categorize them as structural (i.e.\ assumptions motivated by conceptual reasons or used to simplify otherwise extremely challenging technical computations) and as motivated by the requirement of matching with classical gravity. 
\subsubsection{Kinematic assumptions}
Kinematic approximations are related to the properties of the specific states we are considering.
\paragraph*{Structural}
\begin{description}
\item[KS1\label{ass:ks1}] $\hspace{-2pt}$($\hspace{-1pt}$\textbf{\itshape{Condensate states}}, pages \pageref{asspage:KS1a}, \pageref{asspage:KS1b}): 
In this paper, following \cite{Gielen:2013naa,Gielen:2014ila,Gielen:2014uga,Oriti:2015qva,Gielen:2016dss,Oriti:2016qtz,Pithis:2016cxg,Pithis:2019tvp}, we only focus on condensate states, defined as in equation \eqref{eqn:coherentstates}. Condensate states are in fact the simplest representative of the class of coarse-grained states which we expect can be associated to emergent continuum geometries \cite{Gielen:2013naa,Gielen:2014ila,Gielen:2014uga,Oriti:2015qva,Gielen:2016dss,Oriti:2016qtz,Pithis:2016cxg,Pithis:2019tvp}, and thus are of primary interest for the extraction of continuum physics from \qg.
\item[KS2\label{ass:ks2}] $\hspace{-2pt}$($\hspace{-1pt}$\textbf{\itshape{Isotropy}}, pages \pageref{asspage:ks2a}, \pageref{asspage:ks2b}): The condensate wavefunction is required to satisfy the isotropy condition \eqref{eqn:isotropycond}. This assumption remarkably simplifies the computational difficulties related to the derivation of emergent collective dynamics from the microscopic one. It will likely need to be relaxed when one is interested in anisotropic perturbations (see e.g.\ \cite{deCesare:2017ynn}).
\item[KS3\label{ass:ks3}] $\hspace{-2pt}$($\hspace{-1pt}$\textbf{\itshape{Peaking}}, pages \pageref{asspage:ks2a}, \pageref{asspage:ks2b}, \pageref{asspage:ks3c}, \pageref{asspage:ks3d}): The condensate wavefunction, following \cite{Marchetti:2020umh,Marchetti:2020qsq}, is assumed to split into a peaking and into a reduced wavefunction (which is assumed not to spoil the overall peaking properties of the condensate wavefunction\footnote{The validity of this condition can be checked after the solutions of the mean-field dynamics are determined.}), with the former depending only on frame variables, see equation \eqref{eqn:wavefunctioncps} and the discussion in Section \ref{sec:condensates}. The use of coherent peaked states allows to concretely implement a notion of relational evolution with respect to the frame scalar field variables, so that their wavefunction represents a distribution of spatial geometries for each point of the physical manifold labelled by the reference frame fields.

The peaking function is taken to be a Gaussian (with a non-trivial phase) with small width. In the case of a single (clock) variable, with the notation used in equation \eqref{eqn:peakingfunction}, this requirement translates in $\epsilon\ll 1$ \cite{Marchetti:2020umh,Marchetti:2020qsq}. In order to avoid large quantum fluctuations, however, $\epsilon$ cannot tend to zero; it needs to be finite and, in particular, as discussed in \cite{Marchetti:2020umh,Marchetti:2020qsq}, it should satisfy $\epsilon\pi_0^2\gg 1$, where $\pi_0$ determines the non-trivial phase of the Gaussian. This guarantees that all quantum fluctuations of observables associated to the clock variable are small in the classical regime \cite{Marchetti:2020qsq}. Analogous assumptions are made for rod variables.  
\end{description}
\paragraph{Motivated by classical matching}
\begin{description}
\item[KC1\label{ass:kc1}] $\hspace{-2pt}$($\hspace{-1pt}$\textbf{\itshape{Complex width}}, pages \pageref{asspage:kc1a}, \pageref{asspage:kc1b}): The width of the Gaussian determining the peaking on rod fields is in general taken to be a complex parameter $\delta^{(j)}$ (for $j=1,2,3$), with a positive real part $\delta^{(j)}_r>0$ and satisfying ${\delta^{(j)}_i}^2>{\delta^{(j)}_r}^2$, where $\delta^{(j)}_i$ is the imaginary part of $\delta^{(j)}$. This last condition is necessary in order to recover an effective Lorentz signature of second-order derivatives with respect to the frame fields (see e.g.\ the discussion below equation \eqref{eqn:pertvolumegeneralform}). 
It is possible that a more detailed study (guided by the underlying discrete gravity interpretation of the \qg dynamics) of the coupling between matter frames and geometry will relate the validity of this condition to the imposition of local Lorentz invariance. 
\item[KC2\label{ass:kc2}] $\hspace{-2pt}$($\hspace{-1pt}$\textbf{\itshape{Rods rotational invariance of peaking function}}, pages \pageref{asspage:kc2a}, \pageref{asspage:kc2b}): The peaking function is assumed to be rotationally symmetric with respect to rods variables, see equation \eqref{eqn:condensatewavefunction}. This implies in particular $\delta^{(j)}\equiv \delta$, $\pi^{(j)}\equiv \pi_x$. This is necessary in order to obtain a Laplace operator with respect to rods variables. 
\item[KC3\label{ass:kc3}] $\hspace{-2pt}$($\hspace{-1pt}$\textbf{\itshape{Peaking on matter \virgolette{momenta}}}, pages \pageref{asspage:kc3a}, \pageref{asspage:kc3b}): We assume the condensate wavefunction to also be peaked in the variable canonically conjugate to $\matterfield$, $\mommatterfield$, see equation \eqref{eqn:finalcondensatewavefunction}. This is necessary in order to recover both the Friedmann equations already at the background level, see Section \ref{subsec:volumeevolution}. As we discuss in  Section \ref{subsec:volumeevolution}, this \virgolette{momentum peaking} may be related to a semi-classicality condition which could be better understood in models which actually include a scalar field with a non-trivial potential, so that at the right-hand-side of the classical Friedmann equation both the scalar field conjugate variables are present. This is a further direction for future work.
\end{description}
\subsubsection{Dynamic assumptions}
Dynamic approximations are instead related to the specific form of the microscopic GFT action, on how the effective equations for background and perturbed quantities are obtained and, finally, on the specific form of the classical system's dynamics.
\paragraph{Structural}
\begin{description}
\item[DS1\label{ass:ds1}] $\hspace{-2pt}$($\hspace{-1pt}$\textbf{\itshape{GFT action and symmetries}}, page \pageref{asspage:ds1}): The form of the GFT action is in general obtained by comparison with the discrete gravity path integral of the corresponding classical system. In particular, this means that symmetries of the classical action are reflected on the GFT action \cite{Oriti:2016qtz,Gielen:2017eco}. As mentioned in the previous section, the mismatch between \gr and effective \qg equations for perturbations of intermediate and small wavelengths may suggest that some further scrutiny into the details of these models (especially regarding the coupling of geometric and matter degrees of freedom) could be important.
\item[DS2\label{ass:ds2}] $\hspace{-2pt}$($\hspace{-1pt}$\textbf{\itshape{Mean-field dynamics}}, page \pageref{asspage:ds2}): The effective dynamics is taken to be well approximated by a mean-field one, obtained by computing the expectation value of quantum equations of motion on the above coherent states \cite{Oriti:2016qtz}, see equation \eqref{eqn:simplestschwinger}. This assumption implies that microscopic quantum fluctuations are completely neglected, which is certainly not the most general situation one can envisage. In particular, this assumption may be critical exactly because we are interested in small, perturbative effects, which may be heavily affected by quantum corrections to the mean-field dynamics. 

The impact of fluctuations on the mean-field theory has already been studied in the Landau-Ginzburg approach, suggesting the validity of mean-field methods for a class of toy models and simple background configurations \cite{Pithis:2018eaq,Marchetti:2020xvf}. Still, at the time of writing there is no conclusive result for realistic geometric models with non-trivial backgrounds as those considered in this paper. 
\item[DS3\label{ass:ds3}] $\hspace{-2pt}$($\hspace{-1pt}$\textbf{\itshape{Negligible interactions}}, page \pageref{asspage:ds3}): Interaction terms in the effective dynamics are assumed to be negligible with respect to kinetic terms. 
At the mean-field level, this approximation can only be satisfied for condensate densities (or, equivalently, average particle numbers) which are not arbitrarily large \cite{Oriti:2016qtz}. 
\item[DS4\label{ass:ds4}] $\hspace{-2pt}$($\hspace{-1pt}$\textbf{\itshape{Classical system}}, page \pageref{asspage:ds4}):  Frame fields are classically assumed to have negligible impact on the energy-momentum budget of the \universe. Besides making these fields behave as \virgolette{frame-like} as possible, this condition allows to define unambiguously perturbative inhomogeneities with respect to the rods fields.
\end{description}
\paragraph{Motivated by classical matching}
\begin{description}
\item[DC1\label{ass:dc1}] $\hspace{-2pt}$($\hspace{-1pt}$\textbf{\itshape{Mesoscopic regime}}, pages \pageref{asspage:dc1a}, \pageref{asspage:dc1b}): The averaged number of particles of the system is taken to be large enough to allow for both a continuum interpretation  of the expectation values of relevant operators and classical behavior, but not too large that interactions are dominating, see above. In this regime one can see that order two or higher moments of the relevant operators are suppressed (essentially by powers of the averaged number of particles) with respect to expectation values, showing that the averaged dynamics captures sufficiently well the physics of the system \cite{Marchetti:2020qsq}. Moreover, from a more technical point of view, in this regime it is possible to decouple and significantly simplify the equations for linear perturbations \eqref{eqn:firstorder}, see the discussion in Section \ref{subsec:bkgpert}. The validity of this approximation depends of course on the form and on the specific values of the coupling constants of the microscopic interactions.
\item[DC2\label{ass:dc2}] $\hspace{-2pt}$($\hspace{-1pt}$\textbf{\itshape{\virgolette{Single-spin} dominance}}, pages \pageref{asspage:dc2a}, \pageref{asspage:dc2b}): We assume that only one quantum label dominates the evolution (represented by $\sgenlabel$ in our notation). This assumption is justified by the fact that the background evolution is exponential for each $\genlabel$, meaning that, if $\mu_\genlabel$ has a maximum $\genlabel_0$ over the range of $\genlabel$, the evolution of macroscopic observables like the volume will be dominated by $\genlabel_0$ (see e.g.\ equation \eqref{eqn:backgrounddynamics}). The validity of this assumption has also been investigated in \cite{Gielen:2016uft}.
\item[DC3\label{ass:dc3}] $\hspace{-2pt}$($\hspace{-1pt}$\textbf{\itshape{Decoupling}}, page \pageref{asspage:dc3}): We assume that the imaginary part of $\alpha^2$ is much smaller than one, see equation \eqref{eqn:smallimarginarypartalpha}. This requirement mildly constrains the parameters of the states we are considering, and guarantees that the averaged equations for the background match GR. Moreover, together with the assumption of working in a mesoscopic regime, it allows for the first order equations to decouple, see again the discussion in Section \ref{subsec:bkgpert}. 
\item[DC4\label{ass:dc4}] $\hspace{-2pt}$($\hspace{-1pt}$\textbf{\itshape{Effective mass dependence on \virgolette{momentum}}}, page \pageref{asspage:dc4}):
We assume that $\mu_\genlabel(\mommatterfield)\propto \mommatterfield$, a condition that turns out to be necessary in order to match GR already at the background level. If the function $\mu_\genlabel(\mommatterfield)$ admits a series expansion in $\mu_\genlabel(\mommatterfield)$ this condition is naturally satisfied when $\mommatterfield$ is small (and the zeroth order term of the expansion identically vanish). This is not expected to hold in general, but notice that this requirement is imposed only at the point $\mommatterfield=\peakvaluemom$ (see Section \ref{subsec:volumeevolution}). Thus, it can be interpreted as the condition that the momentum of the matter field is not too large (the connection between $\peakvaluemom$ and the classical momentum of the matter field being established in \eqref{eqn:scalarfieldmomentumbackground}). This is expected to be physically well motivated, since we are interested in a semi-classical regime. 
\end{description}

\acknowledgments

We are extremely grateful to Ed Wilson-Ewing for extended discussions and several suggestions, and to Steffen Gielen for important critical remarks and more valuable inputs. We also thank Jean-Luc Lehners for useful comments about cosmological perturbations and gauge choices. LM thanks the University of Pisa and the INFN (section
of Pisa) for financial support, and the Ludwig Maximilians-Universit\"at (LMU) Munich for
the hospitality.

\appendix
\section{Perturbation theory in harmonic gauge}\label{app:harmonicgauge}
Consider four independent minimally coupled scalar fields, let us call them $\framefield^\mu$. They satisfy the massless Klein-Gordon equation $\nabla^a\nabla_a\framefield^\mu=0$. Now, let us introduce coordinates which are \virgolette{adapted} to these scalar fields, let us call them $x^\mu$: $\framefield^\mu\equiv \kappa^\mu x^\mu$, with $\kappa^\mu$ arbitrary constants (no sum over $\mu$). These coordinates also satisfy $\nabla^a\nabla_a x^\mu=0$, which in turn means that, in the coordinate system defined by $x^\mu$, we have
\begin{equation}\label{eqn:harmonicgauge}
   \Gamma^\mu\equiv \Gamma^\mu_{\rho\wfunction}g^{\rho\wfunction}=0\,,
\end{equation}
which is the so called harmonic gauge condition. The coordinates $x^\mu$ are similarly called harmonic coordinates (see \cite{Kuchar1991} for a pioneering work on the relational interpretation of harmonic coordinates and \cite{Marchetti:2020umh, Oriti:2016qtz, Gielen:2018xph, Gielen2020, Gerhardt:2018byq} for applications of harmonic coordinates in the \GFT context).
\subsection{Harmonic condition}
Let us now explicitly write the form of the line element in the harmonic gauge, for a perturbed FRW spacetime.
\paragraph{Background.}
At the background level, we can write the line element as\footnote{We have chosen to denote the lapse function as $N$ to maintain consistency with classical literature. This should not be confused with the the expectation value of the number operator employed in the rest of the paper. In fact, no such confusion should arise, since the number operator and the lapse function are defined only at the quantum and classical level respectively.} ($t\equiv x^0$)
\begin{equation*}
    \diff s^2=-N^2(t)\diff t^2+a^2(t)\delta_{ij}\diff x^i\diff x^j\,.
\end{equation*}
The harmonic gauge condition \eqref{eqn:harmonicgauge} then imposes
\begin{equation*}
    0=\log(N/a^3)'\quad\longrightarrow\quad N/a^3\equiv \clockmomclassic=\text{const.}\,,
\end{equation*}
where a prime denotes a derivative with respect to $t$, so without loss of generality we can fix $\clockmomclassic=1$ and write the background line element as
\begin{equation}\label{eqn:backgroundlineelement}
    \diff s^2=-a^6(t)\diff t^2+a^2(t)\delta_{ij}\diff x^i\diff x^j\,.
\end{equation}
\paragraph{Linear order.}
At the linear order, following \cite{Battarra2014}, we can write the line element in harmonic gauge as
\begin{equation}\label{eqn:perturbedlineelement}
\diff s ^2=-a^6(t)\left(1+2A\right)\diff t^2+2a^4B_{,i}\diff t\diff x^i+a^2(t)\left[(1-2\psi)\delta_{ij}+2E_{,ij}\right]\diff x^i\diff x^j\,,
\end{equation}
where only scalar perturbations have been considered, since they are the only ones of interest for these paper. The first order harmonic gauge condition $\delta \Gamma^\mu=0$ can then be shown to be equivalent to $c^\mu=0$, with \cite{Battarra2014}
\begin{equation*}
    c^0=A'+3\psi'-\nabla^2(E'-a^2B)\,,\qquad c^i=\left[(a^2B)'+a^4(A-\psi-\nabla^2E)\right]_{,i}\,.
\end{equation*}
Therefore, the coordinates $x^\mu=(t,x^i)$ still satisfy the harmonic gauge condition provided that
\begin{subequations}\label{eqn:harmonicgaugeconditionsfirstorder}
\begin{align}
0&=A'+3\psi'+k ^2(E'-a^2B)\,,\\ 0&=(a^2B)'+a^4(A-\psi+k ^2E)\,,
\end{align}
\end{subequations}
where we have moved to Fourier space. It is interesting to notice that these conditions do not completely fix the gauge. Indeed, under an infinitesimal coordinate transformation $x^\mu\to x^\mu+\xi^\mu$, with $\xi^\mu\equiv (\xi^0,\xi^{,i})$, with $\xi^0$ and $\xi$ satisfying 
\begin{equation}\label{eqn:residualgaugefreedom}
    (\xi^0)''+a^4k ^2\xi^0=0=\xi''+a^4k ^2\xi\,,
\end{equation}
the unperturbed metric is still harmonic \cite{Battarra2014}. Perturbations transform correspondingly as
\begin{subequations}
\begin{align}
    A&\to A-(\xi^0)'-3\mathcal{H}\xi^0\,,\\
    B&\to B+a^2\xi^0-a^{-2}\xi'\,,\\
    \psi&\to \psi+\mathcal{H}\xi^0\,,\label{eqn:transfpsi}\\
    E&\to E-\xi\,,\label{eqn:transfe}
\end{align}
\end{subequations}
where $\mathcal{H}\equiv a'/a$  \cite{Battarra2014}. As we will see below, this residual gauge freedom may play an important role in the determination of the perturbations.
\subsection{Dynamics}
Let us now determine the evolution of the quantities entering in the metric. From now on we will neglect the contribution of the reference field to the matter content of the \universe. The system therefore reduces to a perturbed massless scalar field minimally coupled to geometry, which we call $\phi(x)$. At the linear order, therefore, $\phi(x)=\bar{\phi}(t)+\delta\phi(x)$. From now on, we will also set $8\pi G=1$. 
\paragraph{Background.}
The background equations are
\begin{equation}
3\mathcal{H}^2=(\bkgmatter')^2/2\,,\qquad \mathcal{H}'=0\,,\qquad \bkgmatter''=0\,,
\end{equation}
where $\bkgmatter$ is the background scalar field. It is useful to recast the above equations in terms of the background volume $\bar{V}=a^3$. Recalling that $\mathcal{H}^2=[\bar{V}'/(3\bar{V})]^2$, we have that the two Friedmann equations read
\begin{equation}\label{eqn:friedmanneq}
    3\left[\bar{V}'/(3\bar{V})\right]^2=\left(\bkgmattermomclassic\right)^2/2\,,\qquad [\bar{V}'/(3\bar{V})]'=0\,,
\end{equation}
where we have used the definition of the scalar field momentum as 
\begin{equation}
    \bkgmattermomclassic=\frac{\bar{V}}{N}\bar{\matterfield}'=\bar{\phi}'\,,
\end{equation}
with our choice of gauge $N=\bar{V}$. Notice, that if we had chosen $\clockmomclassic\neq 1$, the above Friedmann equations would read
\begin{equation}
     6\left[\frac{\bar{V}'}{3\bar{V}}\right]^2=\left[\frac{\bkgmattermomclassic}{\clockmomclassic}\right]^2\,,\qquad \left[\frac{\bar{V}'}{3\bar{V})}\right]'=0\,,
\end{equation}
\paragraph{Linear order.}
At the linear order, the Einstein equations and the scalar field equations can be reduced to the following system of equations in Fourier space \cite{Battarra2014}:
\begin{subequations}
\begin{align}\label{eqn:bharmonicgauge}
0&=\frac{1}{2}\bkgmatter'\delta\phi'+3\mathcal{H}\psi'+k ^2a^4\psi+k ^2\mathcal{H}(E'-a^2B)\,,\\\label{eqn:secondequationharmonic}
0&=\mathcal{H}A+\psi'-\frac{1}{2}\bkgmatter'\delta\phi\,,\\
0&=E''+k ^2a^4E\,.\label{eqn:eharmonicgauge}
\end{align}
\end{subequations}
Given the very simple matter content that we are considering here, it is easy to obtain simpler equations for the metric perturbations from the above equations and the two gauge conditions \eqref{eqn:harmonicgaugeconditionsfirstorder}. We first combine the two gauge conditions \eqref{eqn:harmonicgaugeconditionsfirstorder} to obtain
\begin{equation}\label{eqn:combinedgauge}
A''+3\psi''+k ^2a^4(A-\psi)=0\,.
\end{equation}
Using this gauge condition in equation \eqref{eqn:bharmonicgauge}, it can be reduced to
\begin{equation*}
\mathcal{H}A'-\frac{1}{2}\bkgmatter'\delta\phi'=k ^2a^4\psi\,.
\end{equation*}
Now, taking a derivative of equation \eqref{eqn:secondequationharmonic}, one finds
\begin{equation*}
\mathcal{H}A'+\psi''-\frac{1}{2}\bkgmatter'\delta\phi'=0\,,
\end{equation*}
where we have used that $\bkgmatter''=0$. Combining these last two equations one then finds
\begin{equation}\label{eqn:psiharmonicgauge}
\psi''+k ^2a^4\psi=0\,,
\end{equation}
which is of the same form of \eqref{eqn:eharmonicgauge}.
The solution to this equation can be expressed in terms of Bessel functions of the first and second kind. Using that $a(t)=a_0\exp[\mathcal{H}t]$, with constant $\mathcal{H}$, we find
\begin{equation}
\psi(t,k)=c_1J_{0}\left(\frac{\sqrt{a_0}k}{\mathcal{H}}e^{2\mathcal{H}t}\right)+2c_2Y_{0}\left(\frac{\sqrt{a_0}k}{\mathcal{H}}e^{2\mathcal{H}t}\right).
\end{equation}
Now, in the limit $k\to 0$, the Bessel function $Y_0$ blows up, so in order to get a finite result we should put $c_{2}=0$. In this case, then, the two relevant super- and sub-horizon limits are
\begin{equation}\label{eqn:limitspsi}
\lim_{k\to 0}\psi(t,k)=c_{1}\,,\qquad \lim_{k\to\infty}\psi(t,k)=0\,.
\end{equation}

From these results one can obtain the evolution equations for all the other relevant quantities. Using the combined gauge condition \eqref{eqn:combinedgauge} one can find that $A$ satisfies the following equation:
\begin{equation}\label{eqn:aharmonicgauge}
A''+a^4k ^2A=4a^4k ^2\psi\,,
\end{equation}
so in the super-horizon limit $k\to 0$ $A$ is forced to be a constant (same as $\psi$) while in the sub-horizon limit $k\to\infty$ $A$ is forced to be equal to $4\psi$ and so it must be zero.

According to these results, we also see from the fact that $\mathcal{H}A+\psi'-\bkgmatter'\delta\phi/2=0$ that in the limit $k\to 0$ $\delta\phi$ is a constant, while in the limit $k\to\infty$ $\delta\phi\to 0$.  This can also be checked explicitly from equation (A.33) of \cite{Battarra2014}, which in our case reads
\begin{equation}\label{eqn:deltaphiharmonicgauge}
\delta\phi''+a^4k ^2\delta\phi=0\,.
\end{equation}

The equation for $B$ can instead be determined from \eqref{eqn:bharmonicgauge}, whose second derivative gives, using equations \eqref{eqn:psiharmonicgauge}, \eqref{eqn:eharmonicgauge} and \eqref{eqn:deltaphiharmonicgauge}:
\begin{equation*}
k ^2\mathcal{H}(a^2B)''=-a^4k ^2\mathcal{H}(a^2B)+a^8k ^4\psi+k ^2(a^4\psi)''\,,
\end{equation*}
which can be more conveniently written as
\begin{equation}
(a^2B)''+a^4k ^2(a^2B)=8a^2(a^2\psi)'\,.
\end{equation}
\paragraph{Perturbed volume equations.}
It is useful to recast the above equations for the metric perturbations in terms of quantities that we have access to from the fundamental quantum gravity theory. The most important one in this context is the local volume element associated to a infinitesimally small patch of spacetime. At the classical level, this can be compared to the local volume element
\begin{equation}
    V_c\equiv \sqrt{\det {_{3}g}}=\sqrt{\det a^2[(1-2\psi)\delta_{ij}+2E_{,ij}]}=a^3\sqrt{\det [\delta_{ij}+2(E_{,ij}-\psi\delta_{ij})]}\,.
\end{equation}
The perturbed part, at first order in $\psi$ and $E$, is therefore given, in Fourer transform, by
\begin{equation}
    \delta V_c=\bar{V}_c(k^2E-3\psi)\,,\qquad \bar{V}_c\equiv a^3
\end{equation}
Since both $E$ and $\psi$ satisfy the same equation, we deduce that 
\begin{equation*}
    (\delta V_c/\bar{V}_c)''+k^2a^4(\delta V_c/\bar{V}_c)=0\,.
\end{equation*}
Using that, by definition, $\mathcal{H}=\bar{V}'/(3\bar{V})$, we find
\begin{equation}\label{eqn:classicalvolumeequation}
    \delta V_c''-6\mathcal{H}\delta V_c'+9\mathcal{H}^2\delta V_c-a^4\nabla^2\delta V_c=0\,.
\end{equation}
In particular, we notice that as a result of \eqref{eqn:limitspsi} (which holds also for the variable $E$, since $E$ and $\psi$ satisfy the same equation), we find that in the super-horizon limit $k\to 0$, we $\delta V_c=\delta V_{c,0}\bar{V}_c$, while in the sub-horizon limit $k\to\infty$, $\delta V_c=0$.
\paragraph{Residual gauge freedom.}
At this point it is interesting to recall that the harmonic gauge condition does not fix entirely the gauge. The residual gauge freedom discussed above allow us to perform an additional gauge transformation $x^\mu\to x^\mu+\xi^\mu$, with the components of $\xi^\mu$ satisfying \eqref{eqn:residualgaugefreedom}. It is interesting to notice that the functional form of the differential equation satisfied by the components of $\xi^\mu$ is the same that the perturbations to the purely spatial part of the metric $\psi$ and $E$ satisfy. In particular, since they transform under this residual gauge freedom as shown in equations \eqref{eqn:transfpsi} and \eqref{eqn:transfe}, this implies that any perturbation in the spatial part of the metric can be reabsorbed by an appropriate gauge choice.
\section{Derivation of reduced wavefunction dynamics}\label{sec:redwfunctiondynamics}
In this section we provide details about the derivation of equation \eqref{eqn:redwfunctionevoprio} from equation \eqref{eqn:fundamentalequationscps}. As mentioned in Section \ref{sec:gftaverageddyna}, the starting point is the expansion of the kinetic term and the reduced wavefunction in power series around $\framefield^0=0$, $\framefieldbf=0$,
\begin{subequations}\label{eqn:expansions}
\begin{align}
\kinetic(\gvariables,h_I;\framefield^{2}_\lambda,\mommatterfield)&=\sum_{\Kexpansindex=0}^\infty \kinetic^{(2\Kexpansindex)}(\gvariables,h_I;\mommatterfield)\frac{(\framefield^2_\lambda)^{\Kexpansindex}}{(2\Kexpansindex)!}\nonumber\\&=\sum_{\Kexpansindex=0}^\infty \frac{\oppkinetic^{(2\Kexpansindex)}(\gvariables,h_I;\mommatterfield)}{(2\Kexpansindex)!}\sum_{\binomexpansindex=0}^\Kexpansindex\binom{\Kexpansindex}{\binomexpansindex}(\framefield^0)^{2(\Kexpansindex-\binomexpansindex)}\vert\framefieldbf\vert^{2\binomexpansindex}(-\lambda)^\binomexpansindex\,,\\
\redwfunction(h_I,\framefield^0+\peakvalue^0,\framefieldbf+\mathbf{x})&=\sum_{\derexpansindextime,\{\derexpansindexspace_{k}\}}\frac{(\framefield^0)^\derexpansindextime}{\derexpansindextime!}\left(\prod_{k=1}^3\frac{(\framefield^i)^{\derexpansindexspace_{k}}}{(\derexpansindexspace_{k})!}\right)\left(\partial_{0}^\derexpansindextime\prod_{k=1}^3(\partial_{k})^{\derexpansindexspace_{k}}\right)\redwfunction(\gvariables,\peakvalue^0,\mathbf{x},\mommatterfield)\,,
\end{align}
\end{subequations}
where $\oppkinetic^{(2s)}\equiv (-\lambda)^s\kinetic^{(2s)}$. As already mentioned in \cite{Marchetti:2020umh}, the expansion of $\kinetic$ in powers of $\chi^2_\lambda$ is not very restrictive in the \GFT context. In fact, it is actually suggested by studies about the inclusion of matter degrees of freedom in \GFT models motivated by lattice gravity considerations \cite{Li:2017uao}. Now we substitute equations \eqref{eqn:expansions} in \eqref{eqn:fundamentalequationscps}, obtaining
\begin{align}
0&=\int\diff h_I \sum_{\Kexpansindex,\derexpansindextime,\{\derexpansindexspace_k\}}\frac{\oppkinetic^{(2\Kexpansindex)}(\gvariables,h_I;\mommatterfield)}{(2\Kexpansindex)!}\sum_{\binomexpansindex=0}^\Kexpansindex\binom{\Kexpansindex}{\binomexpansindex}(-\lambda)^\binomexpansindex\left(\partial_{0}^\derexpansindextime\prod_{k=1}^3(\partial_{k})^{\derexpansindexspace_{k}}\right)\redwfunction(h_I,\peakvalue^0,\mathbf{x},\mommatterfield)\nonumber\\
&\quad\times\left[\int\diff \framefield^0\peakfunc_\cpeakwidth(\framefield^0;\cpeakphase)(\framefield^0)^{2(\Kexpansindex-\binomexpansindex)}\frac{(\framefield^0)^\derexpansindextime}{\derexpansindextime!}\right]\left[\int\diff^3\framefieldbf\peakfunc_{\rpeakwidth}(\vert\framefieldbf\vert;\rpeakphase)\vert\framefieldbf\vert^{2\binomexpansindex}\left(\prod_{k=1}^3\frac{(\framefield^i)^{\derexpansindexspace_{k}}}{(\derexpansindexspace_{k})!}\right)\right]\,.
\end{align}
The integration over $\framefield^0$ gives
\begin{equation}\label{eqn:timeintegral}
I_{2(\Kexpansindex-\binomexpansindex)+\derexpansindextime}(\cpeakphase,\cpeakwidth)=\normalcoeff_{\cpeakwidth}\sqrt{2\pi\cpeakwidth}\left(i\sqrt{\frac{\cpeakwidth}{2}}\right)^{2(\Kexpansindex-\binomexpansindex)+\derexpansindextime}e^{-\cpeakphase^2\cpeakwidth/2}H_{2(\Kexpansindex-\binomexpansindex)+\derexpansindextime}\left(\sqrt{\frac{\cpeakwidth}{2}}\cpeakphase\right)\,,
\end{equation}
where $H_{2(\Kexpansindex-\binomexpansindex)+\derexpansindextime}$ are Hermite polynomials of order $2(\Kexpansindex-\binomexpansindex)+\derexpansindextime$, 
while the integration over $\framefield^i$ is best performed in spherical coordinates. Defining
\begin{equation}
\framefield^3=\radius\cos\pangle\,,\qquad \framefield^2=\radius\sin\pangle\cos\oangle\,,\qquad \framefield^1=\radius\sin\pangle\sin\oangle\,,
\end{equation}
and noticing that the above integral is identically zero for odd $n_{k}$, the integral becomes
\begin{align}
\int\diff^3\framefieldbf\peakfunc_{\rpeakwidth}(\vert\framefieldbf\vert;\rpeakphase)\vert\framefieldbf\vert^{2\binomexpansindex}\left(\prod_{k=1}^3\frac{(\framefield^i)^{2\derexpansindexspace_{k}}}{(2\derexpansindexspace_{k})!}\right)&=\frac{1}{(2\derexpansindexspace_{1})!(2\derexpansindexspace_{2})!(2\derexpansindexspace_{3})!}\int_{0}^\infty\diff \radius \radius^{2(1+b+\derexpansindexspace_{1}+\derexpansindexspace_{2}+\derexpansindexspace_{3})}\peakfunc_{\rpeakwidth}(\radius;\rpeakphase)\nonumber\\
&\quad\times \int_{-1}^1\diff\cos\pangle\,\cos^{2\derexpansindexspace_{3}}\pangle(1-\cos^2\pangle)^{(\derexpansindexspace_{1}+\derexpansindexspace_{2})}\nonumber\\
&\quad\times \int_{0}^{2\pi}\diff\varphi\sin^{2\derexpansindexspace_{1}}\varphi\cos^{2\derexpansindexspace_{2}}\varphi\,.
\end{align}
We compute the three integrals separately. The radial integral
\begin{subequations}
\begin{align}
\radiusint_{\binomexpansindex;\derexpansindexspace_{1},\derexpansindexspace_{2},\derexpansindexspace_{3}}(\rpeakphase,\rpeakwidth)&\equiv \int_{0}^\infty\diff \radius \radius^{2(1+\binomexpansindex+\derexpansindexspace_{1}+\derexpansindexspace_{2}+\derexpansindexspace_{3})}\peakfunc_{\rpeakwidth}(\radius;\rpeakphase)\nonumber\\
&=\normalcoeff_{\rpeakwidth}\int_{0}^\infty\diff \radius \radius^{2(1+\binomexpansindex+\derexpansindexspace_{1}+\derexpansindexspace_{2}+\derexpansindexspace_{3})}e^{-\radius^2/(2\rpeakwidth)}e^{i\rpeakphase \radius}
\end{align}
can be expressed in terms of hypergeometric functions (but the explicit form is not illuminating), while for the angular integrals we have
\begin{align}
\polarint_{\derexpansindexspace_{1},\derexpansindexspace_{2},\derexpansindexspace_{3}}&\equiv\int_{-1}^1\diff\cos\theta\,\cos^{2\derexpansindexspace_{3}}\theta(1-\cos^2\theta)^{(\derexpansindexspace_{1}+\derexpansindexspace_{2})}\nonumber\\
&=(\derexpansindexspace_{1}+\derexpansindexspace_{2})!\Gamma(\derexpansindexspace_{3}+1/2)/\Gamma(3/2+\derexpansindexspace_{1}+\derexpansindexspace_{2}+\derexpansindexspace_{3})\,,\\
\angularint_{\derexpansindexspace_{1},\derexpansindexspace_{2}}&\equiv\int_{0}^{2\pi}\diff\varphi\sin^{2\derexpansindexspace_{1}}\varphi\cos^{2\derexpansindexspace_{2}}\varphi\nonumber\\
&=2\Gamma(\derexpansindexspace_{1}+1/2)\Gamma(\derexpansindexspace_{2}+1/2)/\Gamma(1+\derexpansindexspace_{1}+\derexpansindexspace_{2})\,,
\end{align}
\end{subequations}
where the integrals are evaluated for $\derexpansindexspace_{i}$ non-negative integers. Substituting these equations into the fundamental equation of motion, we have
\begin{align}
0&=\int\diff h_I\sum_{\Kexpansindex,\derexpansindextime,\{\derexpansindexspace_{k}\}}\frac{\oppkinetic^{(2\Kexpansindex)}(\gvariables,h_I;\mommatterfield)}{(2\Kexpansindex)!}\sum_{\binomexpansindex=0}^s\binom{s}{\binomexpansindex}(-\lambda)^\binomexpansindex \left(\partial_{0}^\derexpansindextime\prod_{k=1}^3(\partial_{k})^{2\derexpansindexspace_{k}}\right)\redwfunction(h_I,\peakvalue^0,\mathbf{\peakvalue},\mommatterfield)\nonumber\\\label{eqn:intermediatestep}
&\quad\times\frac{I_{2(\Kexpansindex-\binomexpansindex)+\derexpansindextime}(\cpeakphase,\cpeakwidth)}{\derexpansindextime!}\frac{\radiusint_{\binomexpansindex;\derexpansindexspace_{1},\derexpansindexspace_{2},\derexpansindexspace_{3}}(\rpeakphase,\rpeakwidth)}{(2\derexpansindexspace_{1})!(2\derexpansindexspace_{2})!(2\derexpansindexspace_{3})!}\polarint_{\derexpansindexspace_{1},\derexpansindexspace_{2},\derexpansindexspace_{3}}\angularint_{\derexpansindexspace_{1},\derexpansindexspace_{2}}\,.
\end{align}
Since we are assuming both $\rpeakwidth$ and $\cpeakwidth$ to be small, we can proceed through an evaluation of the lowest order contributions. We have
\begin{align*}
\frac{\radiusint_{0;0,0,0}(\rpeakphase,\rpeakwidth)}{\mathcal{N}_{\rpeakwidth}}&=i\rpeakphase\rpeakwidth^2-ie^{-\rpeakphase^2\rpeakwidth/2}\sqrt{\frac{\pi}{2}}\rpeakwidth^{3/2}(-1+\rpeakphase^2\rpeakwidth)\left(-i+\text{Erfi}(\rpeakphase\sqrt{\rpeakwidth/2})\right),\\
\frac{\radiusint_{1;0,0,0}(\rpeakphase,\rpeakwidth)}{\mathcal{N}_{\rpeakwidth}}&=-i\rpeakphase\rpeakwidth^3\left(-5+\rpeakphase^2\rpeakwidth\right)+e^{-\rpeakphase^2\rpeakwidth/2}\sqrt{\frac{\pi}{2}}\rpeakwidth^{5/2}\left(3+\rpeakphase^2\rpeakwidth(-6+\rpeakphase^2\rpeakwidth)\right)\left(1+i\text{Erfi}(\rpeakphase\sqrt{\rpeakwidth/2})\right),\\
\radiusint_{0;1,0,0}(\rpeakphase,\rpeakwidth)&=\radiusint_{0;0,1,0}(\rpeakphase,\rpeakwidth)=\radiusint_{0;0,0,1}(\rpeakphase,\rpeakwidth)=\radiusint_{1;0,0,0}(\rpeakphase,\rpeakwidth)\,.
\end{align*}
For the angular part, instead, we find,
\begin{equation*}
\polarint_{1,0,0}=\polarint_{0,1,0}=\frac{4}{3}\,,\qquad \polarint_{0,0,1}=\frac{2}{3}\,,
\end{equation*}
and
\begin{equation*}
\angularint_{1,0}=\angularint_{0,1}=\pi\,,\qquad \angularint_{0,0}=2\pi\,.
\end{equation*}
It is interesting to take a look at the above form of the $r$-integrals in the limit of large\footnote{The expansion of the Erfi function is taken in the complex plane, so it is still true as long as the quantity
\begin{equation*}
\vert \combinedpeakparrods\vert=\vert\sqrt{\rpeakphase^2\rpeakwidth/2}\vert=\sqrt{\vert \rpeakphase^2\rpeakwidth/2\vert}=\left(\rpeakphase^2\sqrt{\rpeakwidth_r^2+\rpeakwidth_i^2}/2\right)^{1/2}
\end{equation*}
is very large, which is for instance the case if either $\rpeakwidth_r\rpeakphase^2$ or $\rpeakwidth_r\rpeakphase^2$ is very large.} $\combinedpeakparrods\equiv \sqrt{\rpeakphase^2\rpeakwidth/2}$. In order to do so, we just need the asymptotic expansion of $\text{Erfi}(\combinedpeakparrods)$, which is given by \cite{weisstein}
\begin{equation}
\text{Erfi}(\combinedpeakparrods)\sim -i+\frac{e^{\combinedpeakparrods^2}}{\sqrt{\pi}\combinedpeakparrods}\left(1+\mathcal{O}(\combinedpeakparrods^{-2})\right).
\end{equation}
Retaining only the leading contributions, we then have
\begin{equation}
\radiusint_{0;000}(\rpeakphase,\rpeakwidth)\sim -2\normalcoeff_{\rpeakwidth}\sqrt{2\pi}\rpeakwidth^{3/2}\combinedpeakparrods^2e^{-\combinedpeakparrods^2}\,,\qquad \radiusint_{1;000}(\rpeakphase,\rpeakwidth)\sim 4\normalcoeff_{\rpeakwidth}\sqrt{2\pi}\rpeakwidth^{5/2}\combinedpeakparrods^4e^{-\combinedpeakparrods^2}\,.
\end{equation}
We can now proceed as done in \cite{Marchetti:2020umh} in the case of the single clock field in order to determine an approximate dynamics by simply truncating the above equation essentially at order $\cpeakwidth$ or $\rpeakwidth$ (we assume $\cpeakwidth$ and $\vert\rpeakwidth\vert$ to be of the same order of magnitude.). As a result we need to consider only $\Kexpansindex$ (and thus $\binomexpansindex$), $\derexpansindexspace_{1},\derexpansindexspace_{2},\derexpansindexspace_{3}$ being either $0$ or $1$, and $\derexpansindextime=0,1,2$. Equation \eqref{eqn:intermediatestep} then becomes
\begin{align*}
&\int\diff h_I\biggl\{\polarint_{0,0,0}\angularint_{0,0}\biggl[\radiusint_{0;0,0,0}\left(I_{0}\oppkinetic^{(0)}(\gvariables,h_I,\mommatterfield)+I_{2}\frac{\oppkinetic^{(2)}(\gvariables,h_I,\mommatterfield)}{2}\right)\\
&\quad\qquad-\lambda\radiusint_{1;0,0,0}I_{0}\frac{\oppkinetic^{(2)}(\gvariables,h_I)}{2}\biggr]\redwfunction(h_I,\peakvalue^\mu,\mommatterfield)\\
&\quad+\oppkinetic^{(0)}(\gvariables,h_I,\mommatterfield)I_{1}\radiusint_{0;0,0,0}\polarint_{0,0,0}\angularint_{0,0}\partial_{0}\redwfunction(h_I,\peakvalue^\mu,\mommatterfield)\\
&\quad+\oppkinetic^{(0)}(\gvariables,h_I,\mommatterfield)\frac{I_{2}}{2}\radiusint_{0;0,0,0}\polarint_{0,0,0}\angularint_{0,0}\partial^2_{0}\redwfunction(h_I,\peakvalue^\mu,\mommatterfield)\\
&\quad+\oppkinetic^{(0)}(\gvariables,h_I,\mommatterfield)I_{0}\frac{\radiusint_{1;0,0,0}}{2}\left[\polarint_{1,0,0}\angularint_{0,1}\left(\partial^2_{1}+\partial^2_2\right)+\polarint_{0,0,1}\angularint_{0,0}\partial^2_3\right]\redwfunction(h_I,\peakvalue^\mu,\mommatterfield)\biggr\}=0\mathperiod
\end{align*}
Using the explicit formulae given above for $\polarint$, $\angularint$ and $\radiusint$, as well as equation \eqref{eqn:timeintegral}, we have, factorizing common terms,
\begin{align*}
0&\simeq -4\pi\normalcoeff_{\cpeakwidth}\normalcoeff_{\rpeakwidth}\sqrt{2\pi\cpeakwidth}\sqrt{2\pi\rpeakwidth}(\rpeakwidth \combinedpeakparrods)^2e^{-\combinedpeakparclock^2-\combinedpeakparrods^2}\\
&\quad\times\int\diff h_I\,\oppkinetic^{(0)}(\gvariables,h_I,\mommatterfield)\biggl\{\left[1-r_{2}(\gvariables,h_I,\mommatterfield)\left(\frac{\cpeakwidth}{4}H_{2}(\combinedpeakparclock)+\lambda\rpeakwidth \combinedpeakparrods^2\right)\right]\redwfunction(h_I,\peakvalue^\mu,\mommatterfield)\\
&\qquad +i\sqrt{\cpeakwidth/2}H_{1}(\combinedpeakparclock)\partial_{0}\redwfunction(h_I,\peakvalue^\mu,\mommatterfield) -\frac{\cpeakwidth}{4}H_{2}(\combinedpeakparclock)\partial_{0}^2\redwfunction(h_I,\peakvalue^\mu,\mommatterfield)\\
&\qquad-\frac{\rpeakwidth \combinedpeakparrods^2}{3}\nabla^2\redwfunction(h_I,\peakvalue^\mu,\mommatterfield)\biggr\}\,,
\end{align*}
where we have defined
\begin{equation}
\kinratio_{\Kexpansindex}\equiv \oppkinetic^{(\Kexpansindex)}/\oppkinetic^{(0)}\,,\qquad \combinedpeakparclock\equiv (\cpeakwidth\cpeakphase^2/2)^{1/2}\,.
\end{equation}
In the limit of large $\combinedpeakparclock\gg 1$ we can approximate again the above equation as
\begin{align}
0&\simeq \int\diff h_I \oppkinetic^{(0,0)}(\gvariables,h_I)\biggl\{\biggl[1-r_{2}(\gvariables,h_I,\mommatterfield)(\cpeakwidth \combinedpeakparclock^2+\lambda\rpeakwidth \combinedpeakparrods^2)\biggr]\redwfunction(h_I,\peakvalue^\mu,\mommatterfield)\nonumber\\
&\quad+i\sqrt{2\cpeakwidth}\combinedpeakparclock\partial_{0}\redwfunction(h_I,\peakvalue^\mu,\mommatterfield)-(\cpeakwidth \combinedpeakparclock^2)\partial_{0}^2\redwfunction(h_I,\peakvalue^\mu,\mommatterfield)-\frac{\rpeakwidth \combinedpeakparrods^2}{3}\nabla^2\redwfunction(h_I,\peakvalue^\mu,\mommatterfield)\biggr\}\,.
\end{align}
Now we assume isotropy of the condensate wavefunction, so that, for diagonal kinetic kernels, the above equation reduces to
\begin{equation}
\partial^2_0\redwfunction_{\spin}(x,\mommatterfield)-i\firstdercoeff\partial_{0}\redwfunction_{\spin}(x,\mommatterfield)-\nondercoeff_{\spin}^2(\mommatterfield)\redwfunction_{\spin}(x,\mommatterfield)+\laplaciancoeff^2\nabla^2\redwfunction_{\spin}(x,\mommatterfield)=0\,,
\end{equation} 
where we have suppressed the index $\mu$ in functions of $x^\mu$ for notational simplicity and where we have defined
\begin{equation*}
\firstdercoeff\equiv \frac{\sqrt{2\cpeakwidth}\combinedpeakparclock}{\cpeakwidth \combinedpeakparclock^2}\,,\qquad \nondercoeff_{\spin}^2\equiv\frac{1}{\cpeakwidth \combinedpeakparclock^2}-r_{\spin;2}(\mommatterfield)\left(1+3\lambda \laplaciancoeff^2\right)\,,\qquad \laplaciancoeff^2\equiv \frac{1}{3}\frac{\rpeakwidth \combinedpeakparrods^2}{\cpeakwidth \combinedpeakparclock^2}\,.
\end{equation*}

%\nocite{*}

\bibliographystyle{jhep}
\bibliography{main.bib}
\end{document}